\documentclass[12pt]{article}
\usepackage{times}
\usepackage{ragged2e}
\usepackage{subcaption}
\usepackage{authblk}
\usepackage{tabularx}
\usepackage{float}
\usepackage[export]{adjustbox}
\usepackage{tcolorbox}
\usepackage{listings}
\usepackage{caption}
\usepackage{amsmath}
\usepackage{graphicx}
\usepackage{xcolor}
\usepackage{color}
\usepackage{colortbl}
\usepackage{xspace}
\usepackage{amsmath}
\usepackage{boxedminipage}
\usepackage{comment}
\usepackage{url}
\usepackage{enumitem}
\usepackage{wrapfig}
\usepackage{mathrsfs}
\usepackage{balance}
\usepackage{algcompatible}
\usepackage{algorithm}
\usepackage{multirow}
\usepackage{tikz}

\usepackage[noend]{algpseudocode}

\usepackage[normalem]{ulem}

\newcommand{\pg}{PostgreSQL\xspace}

\newcommand{\CPR}{Cost Productive Rewrite\xspace}
\newcommand{\cpr}{CPR\xspace}
\newcommand{\cprs}{CPRs\xspace}
\newcommand{\qu}{$Q^U$\xspace}
\newcommand{\qt}{$Q^T$\xspace}

\newcommand{\csgm}{CSGM\xspace}
\newcommand{\tsgm}{TSGM\xspace}

\newcommand{\myparagraph}[1]{\noindent \textbf{#1}}

{\begin{list}{$\bullet$}{%
			\setlength{\labelsep}{2pt}\setlength{\leftmargin}{0pt}%
			\setlength{\labelwidth}{0pt}%
			\setlength{\listparindent}{0pt}}}
	{\end{list}}
\newcommand{\denselist}{\itemsep 0pt\parsep=1pt\partopsep 0pt}

\DeclareMathOperator*{\argmax}{arg\,max}

\newcommand{\llmrsq}{LLM-R$^2$\xspace}
\newcommand{\lithe}{{\tt LITHE}\xspace}
\newcommand{\llama}{LLaMA\xspace}
\newcommand{\gpt}{GPT-4o\xspace}
\newcommand{\sota}{{\tt SOTA}\xspace}

\newcommand{\gmSotaMicroDS}{2.49\xspace}
\newcommand{\SotaRewriteMicroDS}{3\xspace}

\newcommand{\gmEnsembleMicroDS}{3.23\xspace}
\newcommand{\EnsembleRewriteMicroDS}{6\xspace}

\newcommand{\gmRedMicroDS}{6.85\xspace}
\newcommand{\RedRewriteMicroDS}{7\xspace}

\newcommand{\gmStatsMicroDS}{10.57\xspace}
\newcommand{\StatsRewriteMicroDS}{9\xspace}

\newcommand{\gmGodMicroDS}{11.84\xspace}

\newcommand{\gmLitheAllDS}{11.5\xspace}

\newcommand{\gmLitheProdArcher}{2.1\xspace}
\newcommand{\LitheProdArcher}{22\xspace}
\newcommand{\gmSotaProdArcher}{1.95\xspace}
\newcommand{\SotaProdArcher}{19\xspace}

\newcommand{\gmLitheProdJOB}{1.9\xspace}
\newcommand{\LitheProdJOB}{4\xspace}
\newcommand{\gmSotaProdJOB}{1.4\xspace}
\newcommand{\SotaProdJOB}{2\xspace}

\newcommand{\gmLitheProdDSB}{7.7\xspace}
\newcommand{\LitheProdDSB}{9\xspace}
\newcommand{\gmSotaProdDSB}{1.7\xspace}
\newcommand{\SotaProdDSB}{3\xspace}

\newcommand{\gmLitheProdStackoverflow}{8.7\xspace}
\newcommand{\LitheProdStackoverflow}{2\xspace}
\newcommand{\gmSotaProdStackoverflow}{7.5\xspace}
\newcommand{\SotaProdStackoverflow}{1\xspace}

\newboolean{showText}
\setboolean{showText}{false}  

\newcommand{\hideEndAlg}[1]{\ifthenelse{\boolean{showText}}{#1}{}}
\makeatletter
\newcommand\email[2][]%
{\newaffiltrue\let\AB@blk@and\AB@pand
	\if\relax#1\relax\def\AB@note{\AB@thenote}\else\def\AB@note{\relax}%
	\setcounter{Maxaffil}{0}\fi
	\begingroup
	\let\protect\@unexpandable@protect
	\def\thanks{\protect\thanks}\def\footnote{\protect\footnote}%
	\@temptokena=\expandafter{\AB@authors}%
	{\def\\{\protect\\\protect\Affilfont}\xdef\AB@temp{#2}}%
	\xdef\AB@authors{\the\@temptokena\AB@las\AB@au@str
		\protect\\[\affilsep]\protect\Affilfont\AB@temp}%
	\gdef\AB@las{}\gdef\AB@au@str{}%
	{\def\\{, \ignorespaces}\xdef\AB@temp{#2}}%
	\@temptokena=\expandafter{\AB@affillist}%
	\xdef\AB@affillist{\the\@temptokena \AB@affilsep
		\AB@affilnote{}\protect\Affilfont\AB@temp}%
	\endgroup
	\let\AB@affilsep\AB@affilsepx
}
\makeatother
            
\begin{document}

\title{Query Rewriting via LLMs}
\date{}

\author[1]{Sriram Dharwada}
\author[1]{Himanshu Devrani}
\author[1]{Jayant Haritsa}
\author[2]{Harish Doraiswamy}

\affil[1]{Indian Institute of Science}
\email{\{\textit{sriramd,himanshud,haritsa}\}\textit{@iisc.ac.in}}
\affil[2]{Microsoft Research India}
\email{\textit{harish.doraiswamy@microsoft.com}}

\maketitle

\pagenumbering{arabic}

\setcounter{page}{1}

\begin{abstract}
When complex SQL queries suffer slow executions despite query optimization, DBAs typically invoke automated query rewriting tools to recommend ``lean'' equivalents that are conducive to faster execution. The rewritings are usually achieved via transformation rules, but these rules are limited in scope and difficult to update in a production system. Recently, LLM-based techniques have also been suggested, but they are prone to semantic and syntactic errors.

We investigate here how the remarkable cognitive capabilities of LLMs can be leveraged for performant query rewriting while incorporating safeguards and optimizations to ensure correctness and efficiency. Our study shows that these goals can be progressively achieved through incorporation of (a) an ensemble suite of basic prompts, (b) database-sensitive prompts via redundancy removal and selectivity-based rewriting rules, and (c) LLM token probability-guided rewrite paths. Further, a suite of logic-based and statistical tools can be used to check for semantic violations in the rewrites prior to DBA consideration. 

We have implemented the above LLM-infused techniques in the \lithe system, and evaluated complex analytic queries from standard benchmarks on contemporary database platforms. The results show significant performance improvements for slow queries, over both SOTA rewriters and the native optimizer. For instance, with TPC-DS on \pg, the GM of runtime speedups was a high \textbf{13.2} over the native optimizer, whereas SOTA only gave \textbf{4.9}.

Overall, \lithe is a promising step toward viable LLM-based advisory tools for ameliorating enterprise query performance.

\end{abstract}


\section{Introduction}
\label{sec:intro}

The SQL queries embedded in enterprise applications are often riddled with inefficiencies and redundancies~\cite{Wetune,stackify_query,bloatquery}, especially when machine-generated by modeling software such as ORM tools (e.g., Entity Framework~\cite{entityframework}, Hibernate~\cite{hibernate}). Query optimizers are expected to, in principle, automatically remove such wasteful bloat while constructing efficient execution plans. However, in practice, they are often led astray by complex representations, resulting in poor response times. 

\begin{figure}[!h]
\centering
\includegraphics[width=0.7\linewidth]{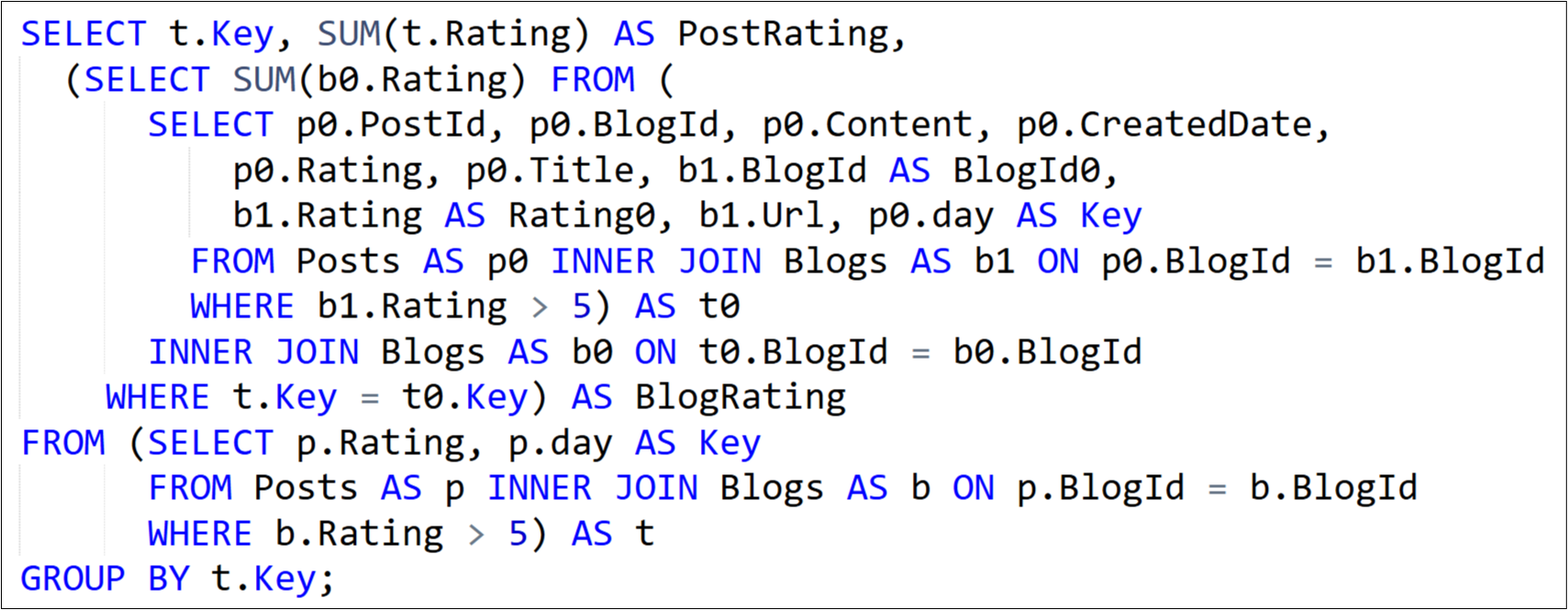}
\vspace{-0.1in}
\caption{Complex SQL Representation}
\label{fig:BloatedQuery}
\vspace{0.15cm}
\includegraphics[width=0.6\linewidth]{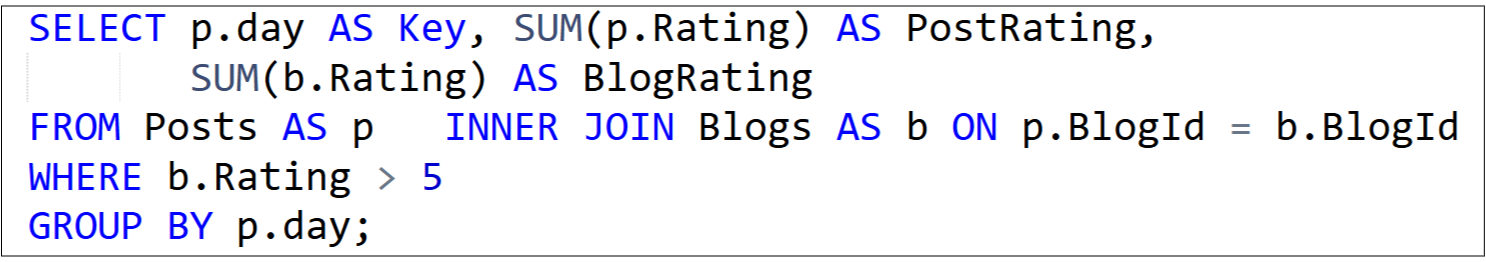}
\vspace{-0.1in}
\caption{Lean Equivalent Query}
\label{fig:LeanQuery}
\vspace{-0.1in}
\end{figure}

As a case in point, consider the blog-processing query~\cite{bloatquery}
shown in Figure~\ref{fig:BloatedQuery},
which computes a daily summary of rating metrics for highly-rated blogs. When this seemingly complex query is given to the current \pg engine (v16)~\cite{pgsql}, its execution plan essentially mimics the hierarchical structure of the query. This strategy leads to multiple scans and joins of the {\tt Blogs} and {Posts} tables,
making the query take several \emph{minutes} to complete.  However, 
the query can be equivalently rewritten (assuming \texttt{NOT NULL} column constraints and key-joins)
in the ``lean'' flat version shown in Figure~\ref{fig:LeanQuery} -- this alternative requires only a single join, and runs more than an \emph{order-of-magnitude} faster, completing within seconds!

For such optimizer-failure scenarios, an alternative option to rectify slow-running queries is to carry out \emph{external tuning} -- in particular, DBAs typically resort to SQL-to-SQL {\bf (S2S)} \emph{query rewriting tools} for recommending performant alternatives.
A viable S2S query transformer should cover a wide range of queries and ensure the rewrite is: (1) semantically equivalent to the original; (2) ideally performance-beneficial, but at least not causing regression; and (3) incurring overheads that are practical for deployment.

\subsection*{Prior Work}
A range of innovative \emph{Rule-based} (e.g.~Learned Rewrite~\cite{Learned_Rewrite}) and \emph{Model-based} (e.g.~Gen-Rewrite~\cite{Genrewrite}) approaches have been proposed for S2S transformations. 
These state-of-the-art (SOTA) techniques have foregrounded the potential benefits of query rewriting. However, as explained later in Section~\ref{sec:Related-Work}, their realization of these benefits is curtailed by: (a) restrictions in rewrite scope, (b) susceptibility to semantic and syntactic errors, and (c) transformations via the plan space (i.e optimization on the nodes of the execution plan tree) rather than directly in query space (i.e. transforming the query structure itself).  In this paper, we address these lacuna and substantively amplify the effectiveness of query rewriting. 

\subsection*{The \lithe Rewriter}
\label{sec:LITHE REWRITE}

\begin{figure}[t]
    \centering
    \includegraphics[width=0.45\linewidth]{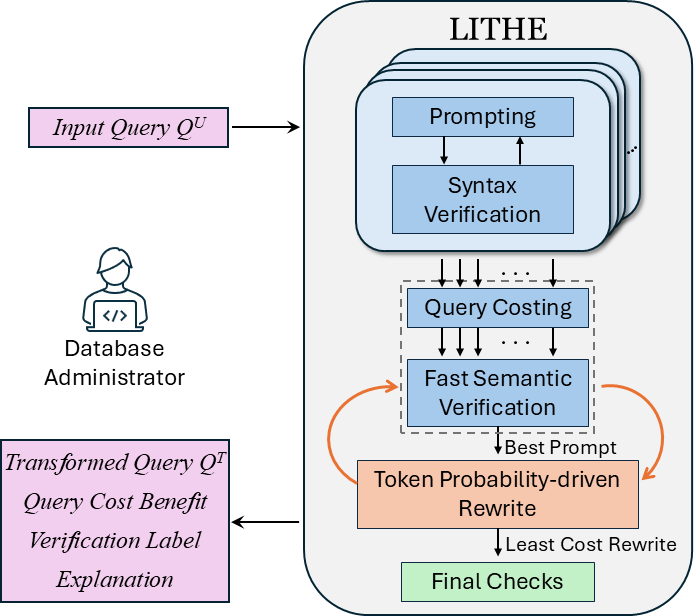}
    \caption{High-level architecture of \lithe}
    \label{fig:lithe}
    \vspace{-0.1in}
\end{figure}

We propose \lithe (LLM Infused Transformations of HEfty queries), an LLM-based query rewriting assistant to aid DBAs in tuning slow-running queries. 
As illustrated in the architectural diagram of Figure~\ref{fig:lithe}, \lithe takes the user query \qu as input and outputs a transformed query \qt, together with 
(a)~the \emph{expected} performance improvement of \qt, in terms of optimizer estimated cost; 
(b)~a \textit{verification label} indicating the semantic equivalence mechanism (provable or statistical) between \qt and \qu; and
(c) a \textit{reasoning} for why the LLM expects \qt to be helpful wrt performance.
Armed with this information, DBAs can leverage their expertise to decide whether or not to use \qt.
Note that having the DBA in the loop is a common practice in commercial query advisory systems~\cite{Dageville2004}.

To help the LLM adapt to different query patterns and structures, we introduce database-aware rules into the prompts, which make use of the rich metadata (e.g., logical schema, selectivities) available in database environments (Section~\ref{sec:dbms-proficient-prompts}). 
Our rules are invoked directly in \emph{query space}, providing the LLM with the latitude to generalize the rule usage to a wide range of queries.
As shown later in our experiments, this is a powerful mechanism for ensuring performant rewrites across database environments.
Furthermore, we leverage the rich telemetry provided by LLMs -- particularly, the  \emph{token probabilities} output at each step in the prediction sequence. Whenever the LLM lacks high confidence in the next token, we follow multiple alternative paths in the decision process. To ensure practical overheads on this enumerative approach, a Monte Carlo tree search (MCTS) technique is incorporated in our implementation (Section~\ref{sec:mtcs-rewrite}).

\lithe also incorporates a suite of logic-based and result equivalence-based semantic equivalence tests to verify the correctness of the rewrites (Section~\ref{sec:sql-equivalence}).
Finally, given that there are often material discrepancies between optimizer cost predictions and actual execution times~\cite{lohmanblog}, \lithe includes heuristic mechanisms, described in Section~\ref{sec:regression}, to identify ``brittle'' rewrites and discard them in favor of the original.

\myparagraph{Results.}
Our experiments on benchmark environments demonstrate that \lithe achieves, for many slow queries, semantically correct transformations that significantly reduce the abstract (optimizer-estimated) costs.
In particular, for TPC-DS, \lithe constructed ``highly productive'' ($ > 1.5x$ estimated speedup) rewrites for as many as 26 queries, whereas SOTA promised such rewrites for only about half the number. Further, the GM (Geometric Mean) of \lithe's cost reductions reached {\bf 11.5}, almost double the {\bf 6.1} offered by SOTA.

We also evaluated whether the projected cost reductions translated into real execution speedups. Here, we find that 
\lithe is indeed often substantively faster at run-time as well.  Specifically, the geometric mean of the runtime speedups for slow queries was as high as \textbf{13.2} over the native optimizer, whereas SOTA delivered \textbf{4.9} in comparison.
Finally, our regression identification techniques were successful in guarding against ``brittle'' rewrites that are promising cost-wise, but eventually poor at run-time.

\subsection*{Contributions}
In summary, our study makes the following contributions: 
\begin{enumerate} 
\item Assesses LLM suitability for S2S transformation. 
\item Transforms directly in query space instead of plan space intermediates, leading to performant rewrites.
\item Incorporates database-sensitive rules in LLM prompts, covering both schematic and statistical dimensions.
\item Leverages LLM token probabilities to guide navigation of the rewrite search space and minimize LLM errors.
\item Evaluates rewrite quality over a range of database environments, demonstrating substantial performance benefits. 
\item Identifies learnings that could help guide research directions for industrial-strength query rewriting.
\end{enumerate}

\section{\lithe Overview}
\label{sec:lithe-architecture}

The \lithe architecture (Figure~\ref{fig:lithe}) comprises a five-stage pipeline:

\myparagraph{1. Prompt-based rewriting.}
This component consists of two modules: LLM Prompting and Syntax Verification.
The user query and an initial prompt are fed to the LLM prompting module, requesting a rewrite. The LLM output's syntax is then verified with the database engine's parser -- if invalid, the error is returned to the LLM via an updated prompt asking for a correction. This goes on iteratively (within a threshold) until a valid SQL query is obtained. 

This process is repeated for each of the different prompts described in Sections~\ref{sec:basic-prompt} and \ref{sec:dbms-proficient-prompts}, and the resulting syntactically-valid rewrites are fed to the query costing component.
We evaluate each prompt within a fresh LLM context, thereby making the prompt processing to be order-independent. 

\myparagraph{2. Query Costing.}
The costs of candidate rewrites are evaluated via the database engine's optimizer. Rewrites whose costs are greater than that of the original query are immediately discarded. Whereas, the potentially beneficial rewrites (if any) are checked for semantic equivalence to the original query.

\myparagraph{3. Fast Semantic Equivalence.}
Statistical (result-based) techniques are employed to quickly and cheaply assess the semantic equivalence of a recommended rewrite.
If the rewrite is deemed valid by this module, it is returned along with the prompt that generated it; otherwise an invalid label is returned.

\myparagraph{4. Token probability-driven rewrite.}
The prompt producing the most beneficial (and valid) rewrite is used as input to a Monte-Carlo tree search (MCTS)-based procedure to further refine the rewrite quality. Specifically, multiple exploration paths are followed whenever the LLM lacks high confidence in the predicted token (details in Section~\ref{sec:mtcs-rewrite}). The query costing and semantic equivalence modules are also used internally within this procedure.

\myparagraph{5. Final Checks and Output.}
The least expensive valid rewrite (as identified by the MCTS module) is tested using a suite of provable (logic-based) techniques to generate a final equivalence label (provable or statistical), and its cost benefit over the original query is computed.
Further, the execution ``brittleness" is assessed using the robustness heuristics. 
If no valid rewrite is identified, or if the rewrite is expected to be a regression, the original query itself is returned to the DBA. Whereas, if a beneficial rewrite is recommended, an LLM-generated reasoning for the expected performance improvement is also extracted. 

\subsection{Performance Framework}

We consider a query that takes more than \emph{T} seconds to complete on the native database engine as a ``slow query'', potentially triggering intervention by the DBA. Based on common industry perception (e.g.~\cite{mysql}), \emph{T} is set at 10 seconds in our study. For this context, we define a \textbf{\CPR (\cpr)} as a rewrite that improves a slow query's performance by at least \textbf{1.5} times wrt the optimizer-estimated cost -- this aggressive choice of threshold is so that: 
(a) there is enough headroom in the optimizer prediction that a runtime regression is unlikely; and (b) the benefits of the rewrite substantively outweigh the transformation overheads.

The overall benefit of a rewriting tool is quantified by the number of \cpr obtained on the slow query workload. Additionally, we measure \textbf{\csgm}, the \textit{geometric mean}~(GM) of the cost speedups obtained by these rewrites. Finally, to assess the actual run-time benefits, we evaluate \textbf{\tsgm}, the geometric mean (GM) of the response-time speedups obtained by these rewrites.
These speedups are measured relative to the original query on the native optimizer.

\subsection{Query Micro-benchmark}
To motivate the progression of strategies incorporated in \lithe, we create an initial micro-benchmark comprising 10 diverse TPC-DS queries for which we were able to \emph{hand-craft} high-quality CPRs. These queries are processed on \gpt, the popular OpenAI LLM, and the rewrites are evaluated on the PostgreSQL v16 database engine. The human rewrites deliver a \csgm of \textbf{\gmGodMicroDS}, serving as an aspirational target to attain computationally.
Later, in Section~\ref{sec:exp}, we extend the evaluation to 
complete benchmarks. 

Further, for ease of presentation, we focus on the \csgm metric in the \lithe design sections (Sections~\ref{sec:basic-prompt},~\ref{sec:dbms-proficient-prompts},~\ref{sec:mtcs-rewrite}). The \tsgm performance is subsequently discussed in Section~\ref{sec:exp}. 

\section{Basic Prompts}
\label{sec:basic-prompt}

\begin{figure}[t]
    \centering
    \includegraphics[width=0.45\linewidth]{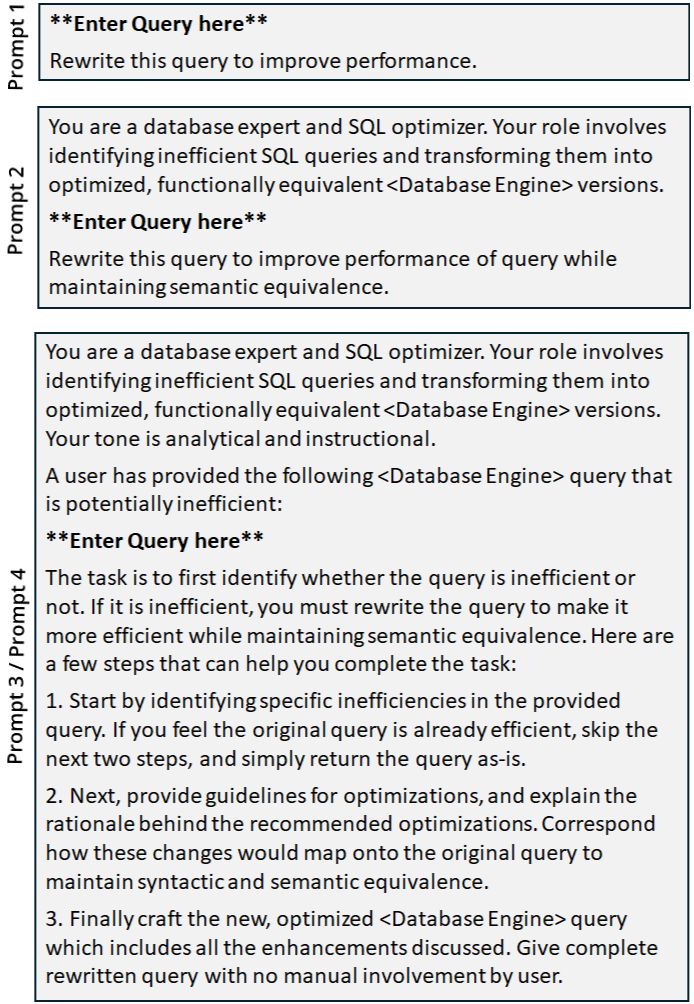}
    \vspace{-0.1cm}
    \caption{Templates used for Basic Prompts}
    \label{fig:baseline-prompts}
    \vspace{-0.2cm}
\end{figure}

In this section, we explore the simplest interface to LLMs, namely \textit{prompting}, for query rewriting.
We evaluate four basic prompts, enumerated in Figure~\ref{fig:baseline-prompts}, which cover a progressive range of instructional detail and test the effectiveness of the LLM's base knowledge. 

\myparagraph{Prompt~1:} 
This is the baseline prompt used in \cite{Genrewrite}, which simply asks the LLM to rewrite a given query to improve performance. 

\myparagraph{Prompt~2:}
Explicit instructions are included to maintain semantic and functional equivalence while rewriting. 

\myparagraph{Prompt~3:}
Verbose instructions are given to rewrite the query, providing step-by-step guidance to the LLM to think rationally. It is first asked to pick out potential inefficiencies in the input query, and then tasked to identify approaches to address these inefficiencies. Finally, it is instructed to apply the identified solution.  Essentially, the prompt tries to make the LLM reason akin to human experts.

\myparagraph{Prompt~4:}
The sequence of instructions in Prompt~3 is split into sub-prompts, and provided to the LLM in an \emph{iterative} manner instead of all at once. This breaks down the complex instructions of Prompt~3 into digestible steps that help the LLM focus on individual tasks.

\subsubsection*{Performance}
Table~\ref{tab:basic-prompt-exp} shows the performance of the four prompt templates on the micro-benchmark. We find that less than half the rewrites are productive with individual prompts. However, a drill-down shows that the best prompt in the ensemble is \emph{query-specific} -- this opens up the possibility of using all four prompts in parallel, and then choosing the best among them. This ensemble approach  raises the \cprs to \textbf{\EnsembleRewriteMicroDS} (Row 5 in Table~\ref{tab:basic-prompt-exp}) -- however, there remain four queries that are not productively rewritten by these prompts. 

\begin{table}[h]
\footnotesize
\centering
\caption{Performance of Basic Prompts.}
\label{tab:basic-prompt-exp}
\vspace{-0.1cm}
\begin{tabular}{|c|c|c|}
\hline
\textbf{Prompt} & \textbf{\# \cpr} & \textbf{\csgm} \\ \hline \hline
Prompt 1 & 3 & 1.99 \\ \hline
Prompt 2 & 2 & 1.85 \\ \hline
Prompt 3 & 4 & 2.83\\ \hline
Prompt 4 & 4 & 3.00 \\ \hline \hline
\textbf{Prompt Ensemble}  & \EnsembleRewriteMicroDS & \gmEnsembleMicroDS \\ \hline \hline
\textbf{SOTA Ensemble} & \SotaRewriteMicroDS & 
\gmSotaMicroDS  \\
\hline
\end{tabular}
\end{table}

The \csgm, shown in the last column of  Table~\ref{tab:basic-prompt-exp}, is at most {\bf 3} for the individual prompts, while the ensemble reaches {\bf \gmEnsembleMicroDS}. But these speedups, although productive, are all lower than those delivered by the human rewrites.

Finally, an ensemble of \sota techniques (described in Section~\ref{sec:exp}) was also processed on the same platform. They delivered  \SotaRewriteMicroDS \cprs with a \csgm of \gmSotaMicroDS (last row in Table~\ref{tab:basic-prompt-exp}), 
indicating the wide gap between the current reality and what is humanly possible.

\section{Database-Sensitive Prompts}
\label{sec:dbms-proficient-prompts}
As discussed above, basic prompting needs to be improved on two fronts: (1) Ensuring productive rewrites where feasible; and (2) Maximizing the impact of these productive rewrites.
Our discussions with industry experts revealed that query inefficiencies are most commonly attributed to redundant operations within the queries themselves. 
In fact, contemporary (non-LLM) optimizers often try to reduce redundancies
during the initial logical optimization phase, prior to physical plan enumeration (e.g., redundant join elimination~\cite{Ahmed2006}).
Furthermore, our analysis of execution plans for
slow queries revealed that the optimizer often fails to substitute filter-related operators, even when beneficial to do so.

To address these issues, we incorporate database domain knowledge into the prompts. Specifically, we design a \textit{one shot}-based prompting template, augmented with a set of database-aware rewrite rules.
As a proof of concept, we explore two categories of rewrites here: (a) Rules that eliminate redundancy in the input queries; and (b) Predicate selectivity-based rules that implicitly guide, via query space reformulations, the query optimizer towards efficient query execution plans.
Of course, this basic set of rules can be expanded further, but as shown by our experiments, even this minimal set is capable of delivering substantive improvements over a broad set of database environments.

The templates used for these prompts (Figure~\ref{fig:rule-prompts}) apply only a \emph{single} rule. This was a conscious design choice because LLMs can be overwhelmed by excessive information given in monolithic form~\cite{2024CHESSpaper}. A related question is whether, while retaining the one-rule-per-prompt design, the rules could be \emph{progressively} applied, thereby benefiting queries with multiple types of redundancies. 
However, the combinatorially large number of rule permutation sequences makes the selection process appear infeasibly expensive. Therefore, we apply each rule using a separate prompt, finally returning the rewrite providing the best improvement. Despite this restrictive policy, highly performant rewrites are obtained in our evaluation (Section~\ref{sec:exp}).

\subsection{Redundancy Removal}
\label{sec:rules-prompt}

There are different types of redundancy that can occur in a SQL query -- repeated computations, superfluous filter predicates, unnecessary joins, etc. Rules {\bf R1 through R4} in Table~\ref{tab:llm-rules} are designed to tackle such redundancies. 
The relevant schematic information (e.g. table names, column names, constraints) required by these rules is also provided in the prompt.

\begin{table}[!h]
\footnotesize
\centering
\caption{Rules for Database-sensitive prompts.}
\label{tab:llm-rules}
\begin{tabular}{|p{0.3cm}|p{12cm}|}
\hline
& \textbf{Redundancy Removal Rules}  \\ \hline
R1 & Use CTEs (Common Table Expressions) to avoid repeated computation of a given expression.\\ \hline
R2 & When multiple subqueries use the same base table, rewrite to scan the base table only once.\\ \hline
R3 & Remove redundant conjunctive filter predicates.\\ \hline
R4 & Remove redundant key (PK-FK) joins.\\ \hline 
& \textbf{Statistics-based Rules} \\ \hline
R5 & Choose EXIST or IN from subquery selectivity (high/low).\\ \hline
R6 & Pre-filter tables involved in self-joins and with low selectivities on their filter and/or join predicates. Remove any redundant filters from the main query. Do not create explicit join statements.\\ \hline
\end{tabular}
\vspace{-0.1in}
\end{table}

\begin{figure}[t]
    \centering
    \includegraphics[width=0.5\linewidth]{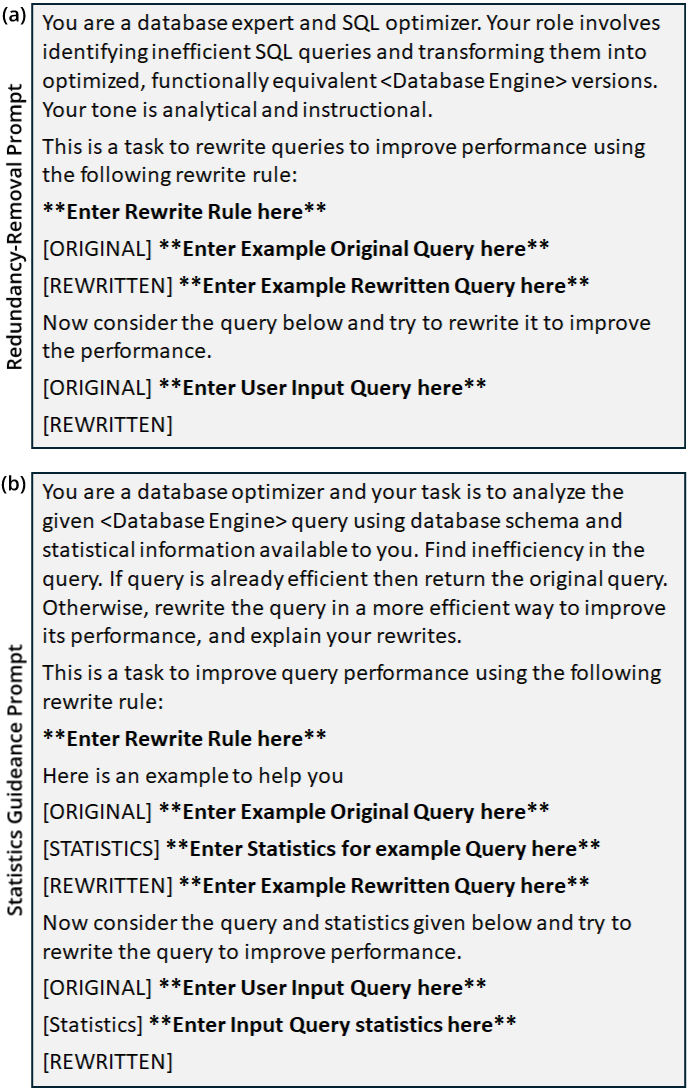}
    \vspace{-0.1cm}
    \caption{Templates for Database-sensitive Prompts}
    \label{fig:rule-prompts}
    \vspace{-0.1cm}
\end{figure}

The template for such rule-based prompts is shown in Figure~\ref{fig:rule-prompts}(a) and includes an \emph{example rewrite} to demonstrate the rule application to the LLM.
Our empirical observation is that the precise wording of the rule instructions is not significant. Instead, what matters are the example rewrites, which ensure the LLM applies the rules effectively, and does not go off on tangential trajectories.
The specific examples used with our rules are listed in Section~\ref{app:rules-ex}. 

\subsubsection*{Performance}
The performance improvement achieved on the micro-benchmark by an ensemble that adds the redundancy-removing prompts to the basic set (Section~\ref{sec:basic-prompt}) is shown in Table~\ref{tab:rule-exp}. We observe that the \cpr increases to \textbf{\RedRewriteMicroDS}, and \csgm grows to \textbf{\gmRedMicroDS}.

\begin{table}[!h]
\footnotesize
\centering
\caption{Performance with Redundancy Removal Rules.}
\label{tab:rule-exp}
\begin{tabular}{|c|c|c|}
\hline
\textbf{Prompt} & \textbf{\# \cpr} & \textbf{\csgm} \\ 
\hline \hline
Basic Prompts $\bigcup$ \{R1, $\ldots$, R4\} & \RedRewriteMicroDS & \gmRedMicroDS  \\ \hline
\end{tabular}
\vspace{-0.1in}
\end{table}

\begin{figure}[t]
    \centering
    \includegraphics[width=0.45\linewidth]{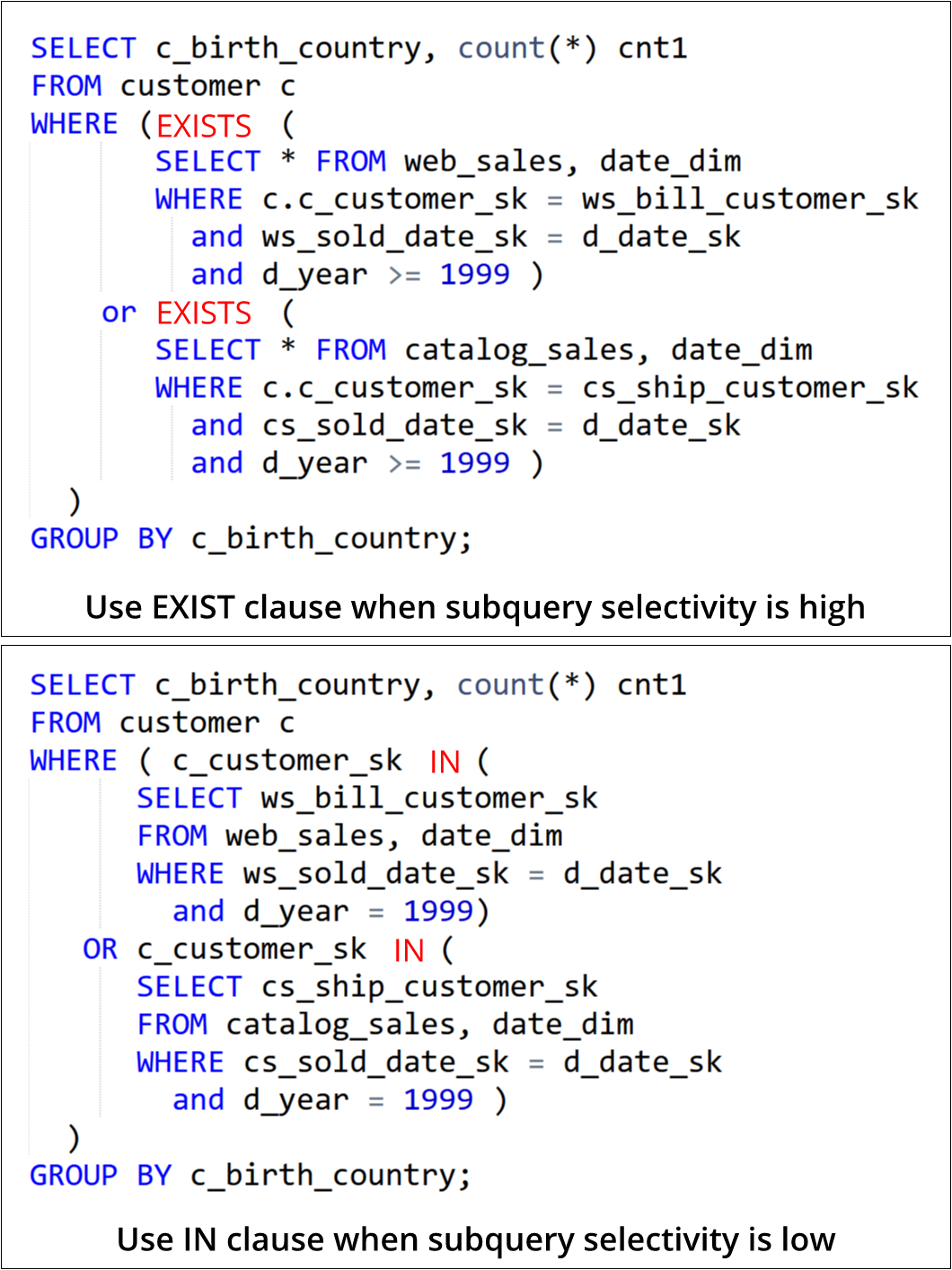}
    \vspace{-0.1cm}
    \caption{Example Queries illustrating Rule~5}
    \label{fig:metadata-example}
    \vspace{-0.1cm}
\end{figure}

\subsection{Selectivity-based Guidance}
\label{sec:stats-prompt}

We now consider rules whose applicability to a query is conditional on the specific database environment, specifically its \emph{statistical} aspects. 
For example, consider the alternative rewrites shown in Figure~\ref{fig:metadata-example} using the EXIST and IN clauses (highlighted in red), respectively. Here the appropriate choice is dictated by the \emph{selectivity} of the inner subquery -- EXISTS for high selectivity values and IN for low values. This observation is encoded in Rule~\textbf{R5} of Table~\ref{tab:llm-rules}. Similarly rule~\textbf{R6}, which pre-filters tables that are involved in self-joins and with low selectivity filter and/or join predicates. 
Note that a specific instruction to not create explicit joins is added in rule R6. This is because in the presence of CTEs, the LLM is prone to schematic confusion regarding which attribute belongs to which table, leading it to construct \emph{invalid joins}.

The input prompt for these rules, as shown in Figure~\ref{fig:rule-prompts}(b), is modified to include the following:
\begin{enumerate}
\item Estimated selectivities of columns appearing in the \texttt{WHERE} and \texttt{JOIN} clauses -- these values are obtained via calls to the cardinality estimation modules of the query optimizer.
\item Clause rewrite rules and instructions based on statistics.
\item Examples relevant to the chosen rewrite rules.
\end{enumerate}

\subsubsection*{Performance}
The performance improvements following addition of selectivity-guided prompts are shown in Table~\ref{tab:metadata-exp}. We observe that \cprs are now obtained for \textbf{\StatsRewriteMicroDS} of the ten micro-benchmark queries. Moreover, the resulting \csgm increases to \textbf{\gmStatsMicroDS}, quite close to the human target of \gmGodMicroDS. 

\vspace{-0.1cm}
\begin{table}[!h]
\footnotesize
\centering
\caption{Performance of Metadata-infused Prompts.}
\label{tab:metadata-exp}
\vspace{-0.1cm}
\begin{tabular}{|c|c|c|}
\hline
\textbf{Prompt}& \textbf{\# \cpr} & \textbf{\csgm}  \\ 
\hline \hline
Basic Prompts $\bigcup$ \{R1, $\ldots$, R6\} & \StatsRewriteMicroDS & \gmStatsMicroDS \\ \hline
\end{tabular}
\vspace{-0.1cm}
\end{table}

We note in closing that rules R1 through R6 not only add queries to the productive category, but also deliver greater improvement for those already deemed to be productive via the standard prompts of Section~\ref{sec:basic-prompt}.
Further, to minimize selection overheads, a classifier could be designed to choose the appropriate rule -- this option is examined in Section~\ref{app:classifier}.

\section{Token Probability Driven Rewrite}
\label{sec:mtcs-rewrite}

A key challenge with LLMs is ``hallucinations'' -- responses that range from being mildly incorrect to completely made up. This is often due to the output tokens having low confidence, which ``confuse" the LLM and generate suboptimal outputs~\cite{hallucinations}.
To have a robust approach for such cases, 
we take inspiration from the code generation literature~\cite{CodeGen}.
Specifically, we propose a 
\textit{Monte Carlo Tree Search}~(MCTS) based decoding approach to search for a sequence of LLM-generated tokens that results in both a valid query rewrite and performance improvements. 
While MCTS has been previously used for \emph{ordering} rule applications in query rewrites~\cite{Learned_Rewrite}, our goal is to guide the LLM in \emph{exploring} these rewrites.

This approach models the problem of query rewriting as a decision tree denoting a \textit{Markov Decision Process (MDP)}~\cite{mdp}. The root node of the tree corresponds to the initial prompt. 
Figure~\ref{fig:mcts} illustrates one such example tree.
An edge from a parent node to a child represents a possible token generated by the LLM and is associated with a value denoting the probability of generating this token given the path taken thus far. This is illustrated by the text (token) and number (probability), on the left and right side respectively, of an edge in Figure~\ref{fig:mcts}.
Here, each edge can be considered as an \textit{action} of the MDP. A node is considered \textit{terminal} if the incoming edge corresponds to the “;” token, signaling end of the textual query.

The \textit{state} of a node $n$ 
is represented by the partial rewrite created by following the path from the root to $n$ -- it is obtained by concatenating the tokens on this path. The root's state is an empty rewrite, and each terminal node's state is a complete rewritten query. 
For example, the state of Node~4 in the example decision tree shown in Figure~\ref{fig:mcts} is the partial rewrite \texttt{``select * from''}.

Given the hundreds of thousands of tokens in LLM vocabularies, it may be very expensive (both financially and computationally) to construct the entire tree. It is therefore essential to significantly reduce the token search space while exploring the tree for valid rewrites.
This is precisely the purpose of MCTS which applies an \textit{Upper Confidence Bound}~(UCB) heuristic~\cite{ucb} to identify the best paths in a tree without computing the entire tree.

\begin{algorithm}[!t]
\caption{Token-augmented Rewrite}
\label{alg:mcts}
\begin{algorithmic}[1]
\REQUIRE 
\Statex $root$ \hspace{0.73cm} \# Start State
\Statex $k$ \hspace{1.1cm} \# Maximum number of child node expansions
\Statex $\theta$ \hspace{1.1cm} \# Probability threshold for node expansion
\Statex $iter_{max}$ \hspace{0.28cm} \# Maximum number of iterations
\smallskip
\State $Potential$, $visits$, $V$ $\gets$ \textbf{empty Map}
\For{$i \gets 1, 2, \dots, iter_{max}$}
    \State $visits[root]$ $\gets$ $visits[root] + 1$
    \State $n_{cur}$ $\gets$ $root$
    \State\textcolor{red}{\# Stage~1: Selection}
    \While{$len(n_{cur}.children) > 0$}
        \State $n_{cur}$ $\gets$ $\argmax_{a \in Actions(n_{cur}.children)}$ \textcolor{blue}{UCB}$(n_{cur}, a)$
        \State $visits[n_{cur}]$ $\gets$ $visits[n_{cur}] + 1$
    \EndWhile
    \State\textcolor{red}{\# Stage~2: Expansion}
    \State $expand \gets$  True
    \While{$expand$ and `;' $\notin n_{cur}.state$}
        \State $tokens_{next}$, $P_{next}$ $\gets$ \textcolor{blue}{Model}($n_{cur},k$)
        \If{$P_{next}[0] \leq \theta$}
            \For{$token \in tokens_{next}$}
                \State $n_{new} \gets$ \textbf{new} $Node$ with \textbf{State} $n_{cur}.state \cdot token$
                \State \textbf{Append} $n_{new}$ to $n_{cur}.children$
            \EndFor
            \State $expand \gets$  False
        \Else
            \State $n_{cur}.state \gets n_{cur}.state \cdot tokens_{next}[0]$
        \EndIf
    \EndWhile
    \State\textcolor{red}{\# Stage~3: Simulation - Expand from $n_{cur}$ to full rewrite}
    \State $query \gets$ \textcolor{blue}{GreedyExpand}($n_{cur}$)
    \State $v \gets$ \textcolor{blue}{ComputePotential}($query$)
    \State $Potential[query] \gets v$
    \State \textcolor{red}{\# Stage~4: Backpropagation}
    \While{$n_{cur} \neq \text{Null}$}
        \State $V[n_{cur}] \gets \max(V[n_{cur}], v)$
        \State $n_{cur} \gets \text{Parent}(n_{cur})$
    \EndWhile
\EndFor
\State\textcolor{red}{\# Return valid rewrite with maximum $Potential > 1$}
\If{$\exists q \in Potential \mid Potential[q] > 1$}
    \State \textbf{return} $q$ having maximum value of $Potential[q]$
\Else
    \State \textbf{return} the original query
\EndIf
\end{algorithmic}
\end{algorithm}

\subsection{MCTS Search Process}
The search procedure is codified in Algorithm~\ref{alg:mcts}, 
and comprises four stages -- \emph{Selection}, \emph{Expansion}, \emph{Simulation} and \emph{Back Propagation} -- described below, which are repeated for $iter_{max}$ iterations.

\myparagraph{1.~Selection:} 
The first stage identifies the most suitable node of the decision tree that is yet to be expanded (i.e., the tokens corresponding to this node have not yet been processed by the search procedure). 
Starting from the root, this process greedily selects successive edges (actions) till an unprocessed non-terminal node is reached (Lines~6--8 in Algorithm~\ref{alg:mcts}). Actions are picked using a UCB that balances exploration and exploitation. The goal is to pick those actions that have either (1)~a higher potential to produce correct and faster rewrites~(exploitation); or (2)~been selected fewer times in the past~(exploration). 

Specifically, given a node $n$ and a set of possible actions $a \in A$, the next node in this traversal is chosen as:
\begin{equation}
n_{\text{next}} = \argmax_{a \in A} UCB(n, a)    
\end{equation}
where $UCB$ is a heuristic sourced from \cite{ucb}, modified to reflect our formulation where values are associated with nodes rather than edges in the tree. It is defined as follows:
\begin{equation}
UCB(n, a) = V(n') + \beta(n) * P_{LLM}(a|n.state) * \frac{\sqrt{log(visits[n])}}{1 + visits[n']}
\end{equation}
Here, $n'$ is the node reached from $n$ by taking action $a$, and the first component $V(n')$ represents the exploitation potential of $n'$ to produce correct and faster queries (this notion is formalized below in Stage~3).
The second component in the equation represents exploration -- it is higher for those child nodes of $n$ that are visited less often. Here, $P_{LLM}$ represents the next token probability,
which can be queried from the LLM by enabling the \texttt{logprobs} option in the API,
and $visits[n]$ is the number of times $n$ has been visited during the search process.
\(\beta\) is a function that controls the balance between exploration and exploitation. It depends on two hyperparameters $c_{base}$ and $c$ -- a higher value of $c_{base}$ makes the algorithm favor exploitation
, whereas a higher value of c increases the incentive to explore. \(\beta\) is defined as:
\begin{equation}
\beta(n) =  log(\frac{visits[n] + c_{base} + 1}{c_{base}}) + c
\end{equation}

\begin{figure}[t]
\small
    \centering
    \includegraphics[width=0.5\linewidth]{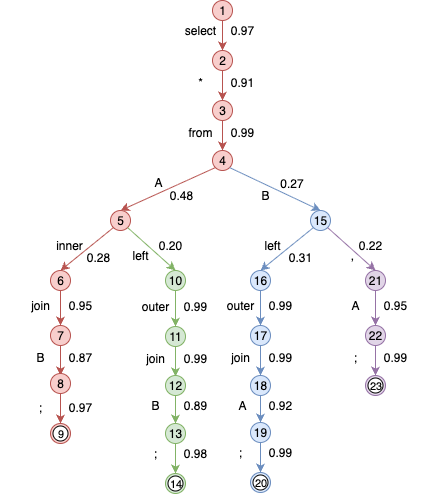}
    \caption{Sample decision tree traversal (Algorithm~\ref{alg:mcts}) 
    }
    \label{fig:mcts}
\end{figure}

For example, consider the state of the tree in Figure~\ref{fig:mcts} when Nodes 5 and 15 are the current unexpanded nodes. At this juncture, the selection procedure will use the UCB values of these two nodes to choose which node to expand next.

\myparagraph{2.~Expansion:} 
The second stage is used to expand the unprocessed node $n_{cur}$ chosen by the Selection stage. Specifically, it retrieves from the LLM the top $k$ probable next tokens from $n_{cur}$'s state (Line~12), and expands the decision tree by adding $k$ new child nodes corresponding to these tokens.
To make the expansion tractable, multiple child nodes are added only if the probability of the highest token falls below a threshold $\theta$ (Line~13). 
Otherwise, the tokens are generated in a greedy fashion from the current node until a point where the LLM is again unsure of the next token, or it reaches a terminal node (i.e. completes a query rewrite).

In our implementation, 
$k$ and $\theta$ are set to 2 and 0.7, respectively. Since branching occurs \emph{only} when this relatively high $\theta$ threshold is breached, most nodes have only \emph{one} child, resulting in long linear sequences. This design, combined with at most two nodes being processed during a branch, significantly reduces the effective fan-out of the tree. Moreover, it ensures that MCTS focuses computational resources on the brittle segments of the rewrite, and does not waste them on robust modifications.
For example, when the root node in Figure~\ref{fig:mcts} is first processed, then Nodes~1, 2, 3 and 4 are expanded and created sequentially since the token probabilities are higher than $\theta$ (Lines~24--26 of the while loop). Only when Node~4 is reached, two new nodes (5 and 15) are created as its children.

\myparagraph{3.~Simulation:} 
The node $n_{cur}$ is expanded greedily based on the highest-probability tokens until a terminal node is reached (Line~21). 
The rewritten query obtained from this greedy expansion is then used to compute the potential $V(n_{cur})$ (Line~22) as follows: for a valid rewrite, $V(n_{cur})$ represents the \textit{speedup} it provides w.r.t. the original query. However, if invalid (i.e. syntactically or semantically incorrect), $V(n_{cur})$ is assigned a zero value. After every simulation, the rewritten query obtained after the greedy expansion of $n_{cur}$ is cached along with $V(n_{cur})$ in a map, $Potential$ (Line~23).

For instance, to compute $V(n_{15})$ in Figure~\ref{fig:mcts}, the blue path in the tree is greedily expanded to identify its rewrite potential. That is, the partial rewrite \noindent ``\texttt{select * from B left outer join A;}'' is used to evaluate $V(n_{15})$.

\myparagraph{4.~Back Propagation:} The $V$ value of the simulation for $n_{cur}$ is back-propagated to all its ancestor nodes, and their values are updated iff the new value is higher than the existing value.

\myparagraph{Rewritten Query.}
At the end of all iterations, $q \in Potential$ with highest value $Potential[q]$ that is greater than 1 is returned as the rewritten query (Lines~29--30).
In case no such rewrite exists, implying that all the valid rewrites are slower than the original query, the original query itself is returned (Line~32).

\subsection{Input Prompt to MCTS}
The root state in Algorithm~\ref{alg:mcts} corresponds to the state just after the prompt is fed to the LLM. 
One way to use this algorithm is to execute it for all the various prompts discussed in the previous sections, and choose the rewrite that provides the best performance. 
This, however, is expensive both from the aspect of query rewrite time, as well as the number of LLM tokens used.
To minimize these costs, given a user query, the \lithe workflow first selects the prompt yielding the most effective rewrite from among the techniques of Sections~\ref{sec:basic-prompt} and \ref{sec:dbms-proficient-prompts}. It then employs this prompt to initiate the MCTS-based rewrite.
In case no prompt provides a lower-cost rewrite, baseline Prompt~1 of Section~\ref{sec:basic-prompt} is used as the fallback option.

\subsection{Performance}

\lithe's performance with MCTS-based rewrites fully matches the human target, yielding \textbf{10} \cprs and the maximum \csgm of \textbf{11.84}. While these gains over just prompting may seem marginal, we wish make two observations: (1)~MCTS extracts maximum improvement whererever possible, and (2)~As shown in Section~\ref{sec:llama}, its impact becomes significant for smaller LLMs like \llama.

\myparagraph{Discussion.}
Our approach implicitly assumes that LLMs \emph{do} possess some degree of knowledge about query optimization -- for example, when presented with a poorly written query versus an optimized one, the LLM is usually capable of identifying why the optimized version would execute better. This expertise arises from the extensive training corpus of LLMs, which includes a large repository of books, papers, and articles on query optimization and execution. 

However, this core knowledge is not fully sufficient to provide good rewrites, and there is a clear need for augmentation with explicit query-specific guidance. This is particularly so when there are multiple plausible optimization paths, or when the LLM may be subtly misled -- this is the reason we leverage MCTS to systematically explore alternatives based on token probabilities. Further, there is explicit evaluation via the database optimizer of the expected improvement in quantitative performance. This separation ensures that our method leverages the LLM's generative capabilities to posit rewrites while relying on an external ground-truth signal (i.e. the database optimizer) to confirm rewrite quality.

\section{Implementation Choices}
\label{sec:impl}

In this section, we briefly discuss the design choices made in our implementation of \lithe.

\subsection{\lithe Parameter Settings}
\label{sec:llm-params}

The \emph{``temperature''} parameter of \gpt, which ranges over [0,1], controls randomness of the model's response. While higher values suit creative tasks that require diverse and exploratory outputs, we seek focused, deterministic answers. Thus, we set the temperature to 0, forcing the model to greedily sample the next token.

The hyperparameters used by \lithe for MCTS are as follows: The maximum number of iterations $iter_{max}$ is set to 8,  expansion threshold $\theta$ is 0.7, and number of expansions $k$ is 2.
The values of $c_{base}$ and $c$ were set to 10 and 4, respectively.
These settings were determined after an empirical evaluation of the various trade-offs, providing a robust balance between efficiency and quality.

Finally, we try a maximum of 5 times to fix, via prompt corrections, any rewrite that exhibits syntax errors (Section~\ref{sec:lithe-architecture}) -- if this threshold is crossed, we drop the rewrite.

\subsection{Query Equivalence Testing}
\label{sec:sql-equivalence}
We use a multi-stage approach, described below, to help the DBA test semantic equivalence between the original query and a recommended rewrite.
As mentioned earlier, having the DBA in the loop is a common practice in commercial query advisory systems~\cite{Dageville2004}.

\myparagraph{1. Result Equivalence via Sampling.}
We use a sampling-based approach to quickly test equivalence in the rewrite generation stages of the pipeline. The idea here is to execute the queries on several small samples of the database and verify equivalence based on the sample results. 
However, while this test is a necessary condition for query equivalence, it is not sufficient. That is,  false positives may be present because the sampled database may not cover all the predicates featured in the query. To minimize this likelihood, we use a combination of (1) \textit{correlated sampling}~\cite{cs2} for maintaining join integrity in the sample, (2) adjusting constants in the filter predicates to produce populated results, and (3) injecting query-specific synthetic tuples in the sample, similar to the XData mutant-killing tool~\cite{xdata}, to cover predicate boundary conditions -- the complete details are in Section~\ref{app:sampling-eq}. 

\myparagraph{2. Logic-based Equivalence.}
Although verifying the equivalence of a general pair of SQL queries is NP-complete~\cite{queryequivalence}, a variety of logic-based tools (e.g. Cosette\cite{Cosette}, SQL-Solver~\cite{SQLSolver}, VeriEQL~\cite{verieql}, QED~\cite{QED}) are available for proving equivalence over restricted classes of queries. 
Once the least-cost rewrite is identified, 
\lithe uses 
QED~\cite{QED} and SQLSolver~\cite{SQLSolver} since these two approaches covered a larger set of queries compared to the alternatives. 
The advantage of such a logic-based approach is that it is definitive in outcome.
Note that this rewrite has already passed the sampling-based tests described above.
 
\myparagraph{3. Result Equivalence on the Entire Database.}
If the logic-based test is inconclusive, result equivalence is evaluated on the entire database itself.
The DBA may choose to prematurely terminate this test in case the checking time is found to be excessive. 

\subsection{Regression Identification}
\label{sec:regression}

We use heuristics to predict whether the actual execution time of a given promising rewrite may turn out to be slower than the original query. The first heuristic is based on the SEER robust plan identification algorithm~\cite{SEER_algorithm}, which is predicated on the database engine providing a ``force-plan'' feature (e.g.~\cite{sqlserver_forceplan}). 
This approach is based on the notion of \emph{replacement safety}, wherein the replacement plan is always either better or within a small sub-optimality factor of the original optimizer-recommended choice. It was
theoretically shown that a safe plan replacement at the \emph{perimeter} of the selectivity space is sufficient to infer safe replacement in the interior of the space as well. The proof used a generic characterization of plan cost functions and the first and second derivatives of these functions at the perimeter. As a matter of efficiency, it was empirically found that instead of the entire perimeter, it was usually sufficient to consider safety just at the \emph{corners}. 

We use the above result to design a regression test as follows:
let $P^U$ and $P^T$ be the plans corresponding to the original query, $Q^U$, and the recommended rewrite, $Q^T$, respectively. We construct \emph{parametrized} versions of these queries, where the constants in the filter predicates are replaced by variables. Then, by assigning appropriate values to these parameters, we construct queries that are located at the \emph{corners} of the selectivity space. The plans $P^U$ and $P^T$ are forced at each of these corner locations, and if the rewrite's cost is lower than the original at \emph{all} of them, the rewrite is deemed to be robust.

At this time, plan forcing is not available in all database engines. For such limited environments, we fall back to an alternative heuristic -- we compare the query runtimes obtained on the series of executions carried out on sampled databases during semantic equivalence testing. If the rewrite's runtimes are lower in \emph{all} of these intermediate executions, then it is estimated to be robust.
We found this extrapolation heuristic to be effective in our experiments due to two reasons: First, the sampled databases, while relatively small compared to the original, are still large in absolute terms (1 GB or more). Second, since most rewrites are related to redundancy removals, a relative performance improvement obtained due to such a rule holds true irrespective of the data sizes.

\section{Experimental Evaluation}
\label{sec:exp}

In this section, we report on \lithe's performance profile. We first describe the experimental setup, including comparative baselines, query suites and evaluation platforms. Then we present the speedup results for both aggregate benchmark and individual queries,  followed by characterization of the rewrite overheads in computational and financial terms. We finally discuss the impact of alternative platforms wrt database engine, database schema and LLMs.

\myparagraph{Rewrite Baselines.}
We compare \lithe with a collection of contemporary rewrite techniques, collectively referred to as {\sota} -- the details of these techniques are provided in Section~\ref{sec:Related-Work}. Specifically, the \sota collection consists of the following approaches:
\begin{enumerate}
\item Baseline LLM prompt~\cite{Genrewrite}: This is Prompt~1 from Section~\ref{sec:basic-prompt}.
\item Learned Rewrite~\cite{Learned_Rewrite}, a purely rule-based rewriter.
\item \llmrsq \cite{LLMR2}, an LLM-guided rule-based rewriter.
\item GenRewrite~\cite{Genrewrite}, a purely LLM-based rewriter.
\end{enumerate}
Given an input query, each of the \sota approaches is independently invoked to perform a rewrite, and the rewrite with the \emph{best} performance is used as the baseline for comparison. 
Note that these approaches may occasionally generate rewrites that are expected by the optimizer's costing module to regress the performance. For safety, we immediately discard such rewrites, similar to \lithe.

\myparagraph{Query Set.}
Our evaluation in this paper primarily focuses on complex analytical queries from the standard TPC-DS decision-support benchmark~\cite{tpcds}, which models a retail supplier environment. The benchmark is used at its default size of 100 GB.
As mentioned earlier, we focus on ``slow queries'' that take over 10 seconds to execute in their original form, operating in a cold-cache environment.

\lithe has also been evaluated on other benchmarks, including DSB~\cite{DSB}, ARCHER~\cite{ARCHER}, JOB~\cite{JOB} and StackOverflow~\cite{Stackoverflow}. The overall performance characteristics were found to be similar to those presented here -- see Section~\ref{app:exp} for details.

\myparagraph{Testbed.}
The experiments were carried out on the following data processing platform: Sandbox server with Intel(R) Xeon(R) CPU E5-1660,
32 GB RAM, and 12 TB HDD, running Ubuntu 22.04 LTS.  A majority of the experiments used PostgreSQL~v16 database engine and \gpt LLM for both \lithe and \sota. Commercial engines and other LLMs are considered in Sections~\ref{sec:commercial-dbms} and ~\ref{sec:altplatform}.

\myparagraph{Metrics.}
For each rewrite technique, we identified the number of queries for which a \cpr (cost productive rewrite with > 1.5 speedup) could be constructed.  Subsequently, we computed the \csgm (Cost Speedup Geometric Mean) and \tsgm (Time Speedup Geometric Mean) performance 
obtained by each technique over the set of all {\cpr}s (i.e. \cprs arising from either \lithe or \sota). Cost speedups are computed relative to the native optimizer, and runtime speedups are measured as the ratio of original query runtime to rewritten query runtime in a cold-cache environment.

From the investment perspective, we measured the average rewrite time per query, and additionally for the LLM-based techniques, the number of tokens used in the rewrite process.

\subsection{Rewrite Quality (Cost and Time)}
\label{sec:rewrite-perf}

\subsubsection{Estimated Cost}
\label{sec:rewrite-cost}

\lithe produces a rewrite with a positive cost speedup (> 1x) for 46 of the 88 TPC-DS queries deemed to be slow by our threshold. Of these 46, there were \textbf{26} \cprs resulting in a \emph{highly productive} \csgm of \textbf{11.5}. On the other hand, \sota delivers only \textbf{13} \cprs (out of 42 positive rewrites) with a \csgm of \textbf{6.1}. All but one of the \sota \cprs also feature in the \lithe \cprs, making the total number of \cprs considered to be 27.  Of these 27, we were able to formally verify 11 using the logic-based tools, whereas the remaining 16 passed our statistical tests. Furthermore, we also manually verified the correctness of these rewritten queries.

\begin{figure}[t]
  \begin{minipage}[t]{.49\textwidth}
    \includegraphics[width=\textwidth]{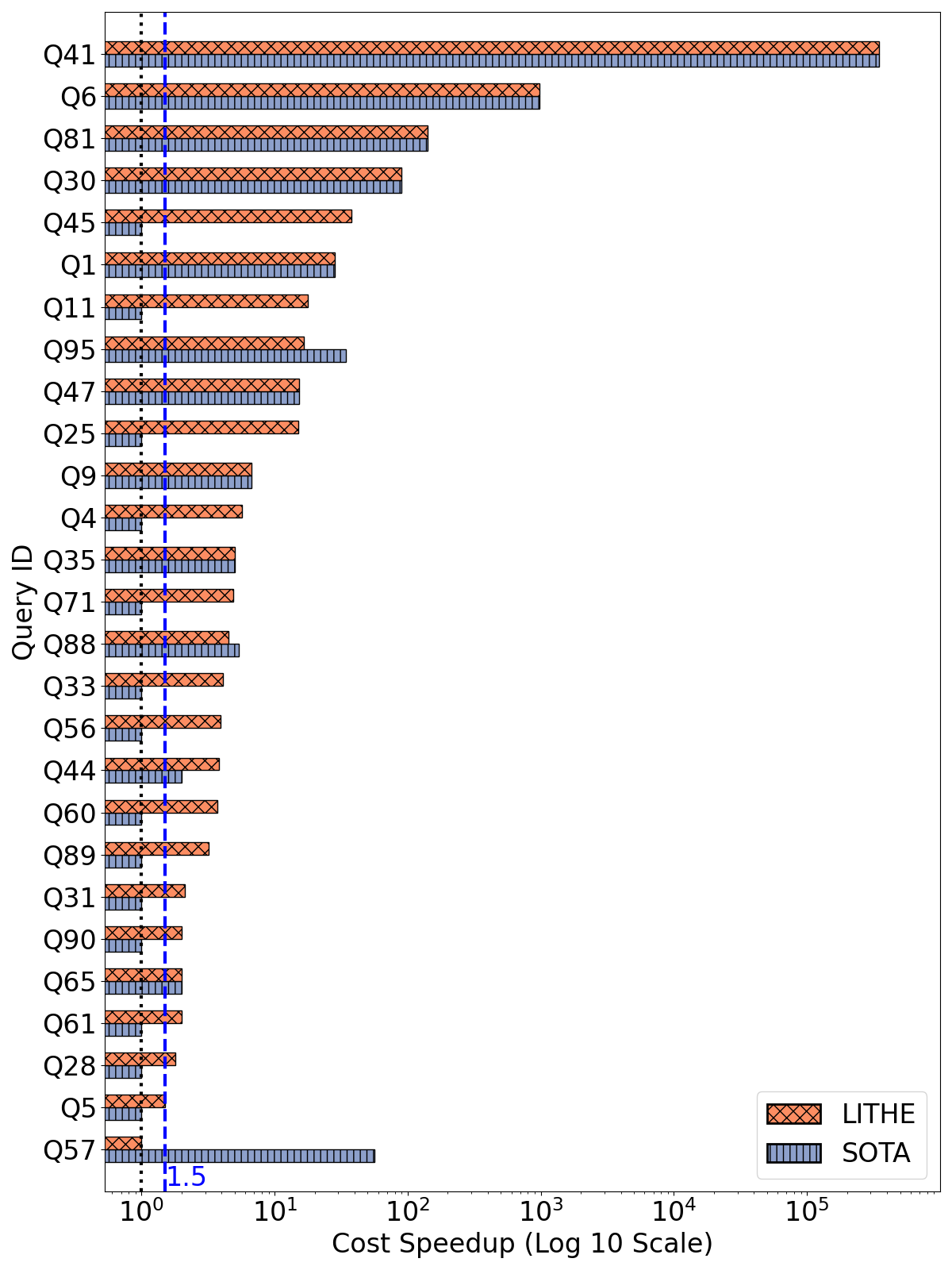}
    \caption{Plan Cost Speedups}
    \label{fig:query-speedup}
  \end{minipage}
  \hfill
  \begin{minipage}[t]{.49\textwidth}
    \includegraphics[width=\textwidth]{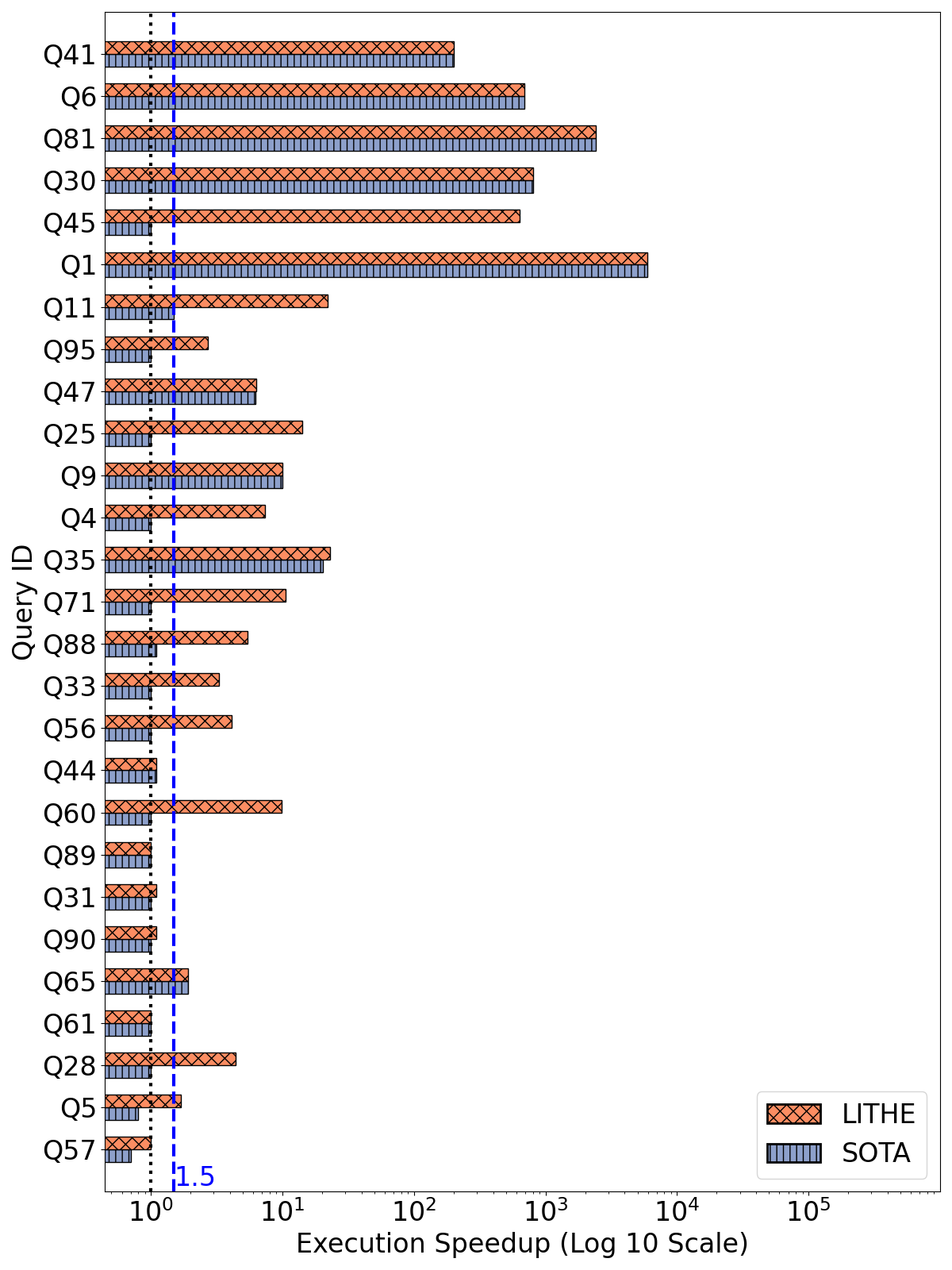}
    \caption{Execution Time Speedups}
    \label{fig:execution-time}
  \end{minipage}
\end{figure}

A drill-down into the cost speedup performance at the granularity of individual queries is shown in Figure~\ref{fig:query-speedup}, which compares \lithe (orange bars) and \sota (blue bars) on each of the 27 \cpr queries  -- note that the cost speedups on the $x$-axis are tabulated on a $\mathbf{log_{10}}$ scale, and the queries are sequenced in decreasing order of \lithe speedup. The vertical dotted line at $1$ represents the normalized baseline cost of the original query with the native optimizer, while the vertical line at $1.5$ is the \cpr threshold.

We first observe, gratifyingly, that rewrites are indeed capable of promising dramatic cost speedups over the native engine -- take, for instance, Q41, which improves by a whopping \emph{five orders-of-magnitude} for both \sota and \lithe. This improvement in query performance is due to replacing the ``\texttt{WHERE (SELECT COUNT(*) from ...) > 0}" clause with ``\texttt{WHERE EXIST (SELECT 1 from ...)}" -- the latter efficiently checks result existence in an inner subquery since it removes the costly aggregation function.

Second, in as many as 15 queries, \lithe's cost speedup \emph{substantively exceeds} \sota, while in the remainder it largely matches \sota. In fact, for several queries (e.g. Q45, Q25, Q4) \lithe produces a highly beneficial \cpr but \sota returns the original query. The only case where \sota is appreciably better is Q57 where it projects a large speedup but \lithe settles for the original query.  And in Q88 and Q95, \sota performs marginally better.

At this stage, one might expect that adding more rules to \lithe could bring it on par with \sota for queries like Q57. However, we deliberately include only broad-brush rules in \lithe to ensure generalizability and efficiency. Moreover, as the following timing section shows, due to this conservative approach \lithe actually \emph{outperforms} \sota on these queries (Q57, Q88, Q95) at runtime!

We also analyzed queries where \lithe was unable to provide CPRs to determine whether there were common patterns that led to unsuccessful outcomes. Our audit indicated that the primary reasons were: (a) Structural Simplicity -- for instance, flat SPJ queries without nesting that are already conducive to current optimizers (e.g. Q3, Q15, Q52), or (b) Structural Tightness -- for instance, queries without repeated computations where the scope for improvement is inherently limited (e.g. Q7, Q19, Q22).
Further, some failures may be due to the \emph{specific} LLM being used, and not the queries themselves (see Section~\ref{sec:llama}). Therefore, one could consider implementing an \emph{ensemble} of diverse LLMs to gain enhanced beneficial coverage.

\subsubsection{Execution Time}
\label{sec:execution-times}
Thus far, we had considered optimizer-estimated costs. We now move to wall-clock runtimes for query executions -- Figure~\ref{fig:execution-time} shows the runtime speedups (on a $\mathbf{log_{10}}$ scale) obtained by \lithe and \sota. 
We again observe that there are indeed several queries where substantial time benefits are achieved by the rewrites over the native engine, even exceeding \emph{order-of-magnitude} benefits in some cases -- for instance, \lithe improves Q45 by a huge factor of 700!
Further, for 12 of the 26 queries, \lithe significantly outperforms {\sota} -- as a case in point, by {\bf 14} times in Q11. In the remainder, it matches \sota, including as mentioned above, queries where \sota's optimizer costs were better.

From a modeling perspective, we see that the well-documented gap between optimizer predictions and actual run-times is prevalent in the rewrite space as well. On the one hand, there is Q45 where the projected speedup of $40$ increases to $700$ at runtime, whereas on the other, the $10^5$ speedup for Q41 decreases to $200$.  But for \sota, the reductions can be severe -- a striking case in point is Q57, where \sota actually causes \emph{regression} despite a speedup projection of close to 100. 
However, the good news is that with \lithe, although the runtime speedups did not always match the projections, regressions were not encountered among the \cpr rewrites thanks to the sampling-based checks.
Overall, \lithe produces more robust rewrites resulting in a high \tsgm of \textbf{13.2}, whereas \sota only delivers \tsgm of \textbf{4.9}.

\subsubsection{Reasoning}
As a confirmatory exercise, we compared the LLM-generated explanation output by \lithe with our own manual analysis of the query plans generated for the original and rewritten queries. 
For example, consider Q45 -- the provided explanation correctly lists the main reasons for the $700$x speedup as ``\textit{the use of CTEs}", ``\textit{Reduction in data volume early on}" due to pre-filtering based on join predicates, etc.
Overall, we found that the explanations provided by \lithe matched our manual analysis of the plans, indicating that model-based reasoning is well aligned with human-backed reasoning in these scenarios.

\subsection{Ablation Analysis of Rewrite Quality}
\label{sec:ablation}

\subsubsection{Components of  {\lithe}}
A natural question at this stage is the role of the various techniques in \lithe towards achieving its large performance benefits. This analysis is captured in Table~\ref{tab:lithe-contribution} which lists the {\cpr}s for each technique when invoked in \emph{isolation} as well as the \emph{cumulative} number of {\cpr}s when the different techniques are combined in the order of listing.

\begin{table}[h]
\footnotesize
\centering
\caption{{\cpr}s contributed by \lithe components.}
\label{tab:lithe-contribution}
\begin{tabular}{|l|c|c|} \hline
 & \multicolumn{2}{c|}{\textbf{\# CPR}} \\
 & \textbf{Isolated} & \textbf{Cumulative}  \\ \hline \hline
Basic Prompts (Section~\ref{sec:basic-prompt}) & 15  & 15  \\ \hline
Rules R1 --- R4 (Section~\ref{sec:rules-prompt}) & 10 & 18  \\ \hline 
Rules R5, R6 (Section~\ref{sec:stats-prompt})    & 11 & 25  \\ \hline 
MCTS (Section~\ref{sec:mtcs-rewrite})            & 26 & 26 \\ \hline 
\end{tabular}
\end{table}

We observe that the Basic Prompt ensemble and Redundancy Rules (R1 -- R4) contribute to around two-thirds of the {\cpr}s (18/26). When the statistics-infused Rules (R5, R6) are added, this number jumps to 25. Finally, MCTS in isolation (with Prompt~1 as the seed prompt) produces only 11 CPRs. However, by using the best prompt as seed, an additional \cpr is obtained and the cost speedup of a pre-existing \cpr is also improved.

\subsubsection{Database-Sensitive Prompts.}
Since MCTS explores the search space at a fine granularity, one could ask whether just the Basic Prompts in conjunction with MCTS would suffice to provide good performance. The motivation is that it would relieve us from using the database-sensitive rules R1--R6
which incur significant computational and financial overheads. When this experiment was conducted, the \csgm dropped precipitously to a paltry \textbf{5}, a far cry from the \gmLitheAllDS obtained with the database-sensitive rules. These results highlight the need to reflect database awareness for effective query rewriting, and not rely solely on prior LLM knowledge.

\subsection{Rewrite Overheads (Time/Money)}
\label{sec:overheads}
Having established the performance benefits of rewrites, we now turn our attention to their time and financial overheads.

\subsubsection{Transformation Time}
We first look at the end-to-end time required to perform a rewrite (i.e. to run the entire pipeline shown in Figure~\ref{fig:lithe}).
Table~\ref{tab:different_benchmark_cost} shows the average processing time per \cpr query, where we see that the \lithe rewrite process takes a few minutes. 
However, note that this investment may be acceptable in deployment given that the execution benefits typically far outweigh the compilation overheads. 
For instance, with Q11, the original query took nearly \emph{an hour} to complete, whereas the \lithe rewrite executed in  under 3~\emph{minutes}.
Further, many applications tend to use a set of canned queries which are run thousands of times. Thus, even a large one-time investment can be easily recovered over repeat executions of such queries.

\begin{table}[h]
\footnotesize
\centering
\caption{Rewrite Overheads of \lithe and \sota.}
\label{tab:different_benchmark_cost}
\begin{tabular}{|c|c|c|c|}
\hline
& \textbf{Avg. Time (min)} & \textbf{Avg. Tokens } & \textbf{Avg. Cost (USD) } \\ \hline \hline
LITHE       & 5  & 18427 & 0.045  \\ \hline
SOTA        & 1.7 & 20076 & 0.050 \\ \hline
\end{tabular}
\end{table}

Notwithstanding the above, we also observe that \lithe is considerably slower than \sota in producing rewrites. A drill-down 
showed the lion's share of the time is taken by the initial prompt ensemble and the final MCTS module. We explore techniques to improve their efficiency in Section~\ref{app:classifier}. 

\subsubsection{Monetary Outlay}
The average number of LLM tokens required by \lithe and \sota, and their associated financial costs\footnote{At the time of writing, \gpt costs USD 2.5 per million tokens.}, are also shown in Table~\ref{tab:different_benchmark_cost}. The good news is that the inference charges per query are just a few cents, making rewriting practical from a deployment perspective.

\subsection{Commercial DBMS}
\label{sec:commercial-dbms}

A legitimate question could be whether the rewrites made amends for the \pg optimizer but may fail to be useful in highly-engineered database engines. To evaluate this issue,
we performed TPC-DS rewrites on a pair of popular commercial DBMS, which we refer to as \textbf{OptA} and \textbf{OptB}. 

\vspace{-0.1cm}
\begin{table}[h]
\footnotesize
\centering
\caption{Performance on Commercial Database Engines.}
\label{tab:commercial-database-experiment}
\vspace{-0.1cm}
\begin{tabular}{|l|c|c|c|c|c|c|}
\hline
& \multicolumn{2}{c|}{\textbf{\# CPR}} & \multicolumn{2}{c|}{\textbf{\csgm}} & \multicolumn{2}{c|}{\textbf{\tsgm}} \\ \hline
& OptA & OptB & OptA & OptB & OptA & OptB \\ \hline
\hline
\lithe & 12 & 9 & 3.6 & 4.1 & 2.1 & 1.9 \\ \hline
SOTA &  3 & 5 & 1.5 & 1.3 & 1.4 & 1.2 \\ \hline
\end{tabular}
\end{table}

The performance of \lithe and \sota on these two systems is shown in Table~\ref{tab:commercial-database-experiment}, with \lithe continuing to do better than \sota. Despite the apparent lack of optimization headroom, \lithe still produces 12 and 9 \cprs resulting in a healthy \csgm of \textbf{3.6} and \textbf{4.1}, respectively. Further, the \tsgm provided by these rewrites are a useful \textbf{2.1} and \textbf{1.9}, respectively.

Interestingly, although we did not observe any regressions with \pg, a few did surface in the commercial systems. Nevertheless, our regression identification mechanisms effectively caught these brittle rewrites. As a case in point, a promising rewrite estimated by one of the commercial optimizers for Q23, actually takes 37 minutes to complete as compared to 18 minutes for the original query -- this doubling slowdown was successfully flagged by the SEER heuristic (Section~\ref{sec:regression}), and the rewrite was abandoned.

The above results suggest \lithe has a useful role to play in industrial environments. From a different perspective, a company building a new database engine could use \lithe to non-invasively overcome the limitations of early versions of its optimizer.

\subsection{Dependence on LLM}
\label{sec:altplatform}

\subsubsection{Impact of Training}
\label{sec:masked}
An interesting question to ask now is whether the performance benefits seen thus far could be an artifact of \gpt having already been trained well on the TPC-DS benchmark, which is  prominent in the public domain.

To investigate this issue, we ran \lithe on two datasets that were confirmed to be \emph{unknown} to the GPT-4o version used in our evaluation. 
The first is the recently released Football benchmark~\cite{footballdb} for Text-to-SQL evaluation. From the query suite, which largely consists of
simple SPJ and Union queries, we selectively picked the few complex queries. Further, we created our own complex queries to include in the test suite. On this enriched workload, \lithe on \pg produced 11 \cprs with \csgm of 12 and \tsgm of 1.5, while SOTA gave 6 \cprs with \csgm of 3 and \tsgm of 1.1.

The second dataset is a proprietary customer benchmark used extensively by an industrial development team. For this environment, \lithe on a commercial engine produced 8 \cprs  with a \csgm of 2 and a \tsgm of 3.3. Note that these improvements are achieved \emph{despite} this engine having been fine-tuned on this workload over an extended period. Moreover, the numbers are consistent with those reported for TPC-DS in Section~\ref{sec:commercial-dbms}.

\subsubsection{Impact of Model}
\label{sec:llama}

In our concluding experiment, we evaluated \lithe's performance on the \llama~3.1 70~billion parameter instruct model, substantially smaller compared to \gpt, which may have several hundred billion parameters~\cite{toi-article}. 
To attain practical inference times, the model was loaded using a low 4-bit quantization. Further, to ensure reproducibility and deterministic answers, the \textit{do\_sample} parameter was set to \textit{False}, which forces the LLM to perform greedy decoding. To make up for the huge reduction in model parameters as compared to \gpt, we include up to two example demonstrations for each rule-based prompt.

For this environment, Table~\ref{tab:mcts-exp-llama} shows \lithe's performance on the TPC-DS queries with and without MCTS. Although certainly lower than the corresponding numbers with \gpt (Section~\ref{sec:rewrite-perf}), it is encouraging to see that, in absolute terms, significant performance benefits can be obtained for most queries, especially with MCTS support. So, the message is that smaller models can also be fruitfully used in real-world environments.

\vspace{-0.1cm}
\begin{table}[!h]
\footnotesize
\centering
\caption{\lithe Rewrite Performance with \llama}
\label{tab:mcts-exp-llama}
\begin{tabular}{|c|c|c|c|}
\hline
& \textbf{\# \cpr} & \textbf{\csgm} & \textbf{\tsgm} \\ \hline \hline
\llama without MCTS & 18 & 5.6 & 6.5 \\ \hline
\llama with MCTS    & 22 & 8.5 & 10.9 \\ \hline
\end{tabular}
\end{table}

Finally, we also evaluated {\lithe} on {\bf Gemini 2.5 Flash}. Here, \lithe produced 17 \cpr with \csgm of 7.2 and \tsgm of {\bf 5.8}. Although not as strong as GPT-4o, these results highlight the potency of \lithe's recommendations across a range of LLM platforms.

\section{The Road Ahead}
\label{sec:ll}

Based on our study, we now present a few observations with implications for the future design and deployment of rewriting tools.

\subsection{Rewrite Space Coverage by LLMs}
Given the decades-long research on database query optimization, we expected the potential for performance improvement via rewriting to be limited. What came as a surprise was the substantial scope for improvement still available, as showcased by the large \csgm and \tsgm values, even on commercial platforms. 
These results suggest that LLMs explore optimization spaces that are well outside the purview of contemporary database engines. 
Specifically, rewrites in the query space appear to guide the optimizer to explore fresh regions of the plan space not evaluated with the original query formulation. 
Further, this enhanced space could be augmented, in a two-stage process, with the recent proposals for LLM-based ``plan hints'' that steer the optimizer in fruitful directions within a plan space~\cite{LLMplanhint}. 
We plan to explore these space relationships and options in our future work.

\subsection{Rewrite Migration to Optimizer}
The above demonstrated the potent exploratory power of LLMs. But from an overheads perspective, such rewrites should ideally be within the optimizer's native search space rather than recommended from outside. Therefore, it would be a useful exercise to try and distill fresh optimization rules from these instances, leveraging the extensibility features of contemporary optimizers~\cite{extoptbook} to facilitate their incorporation in existing systems.

\begin{figure}[h]
    \centering
    \includegraphics[width=0.5\linewidth]{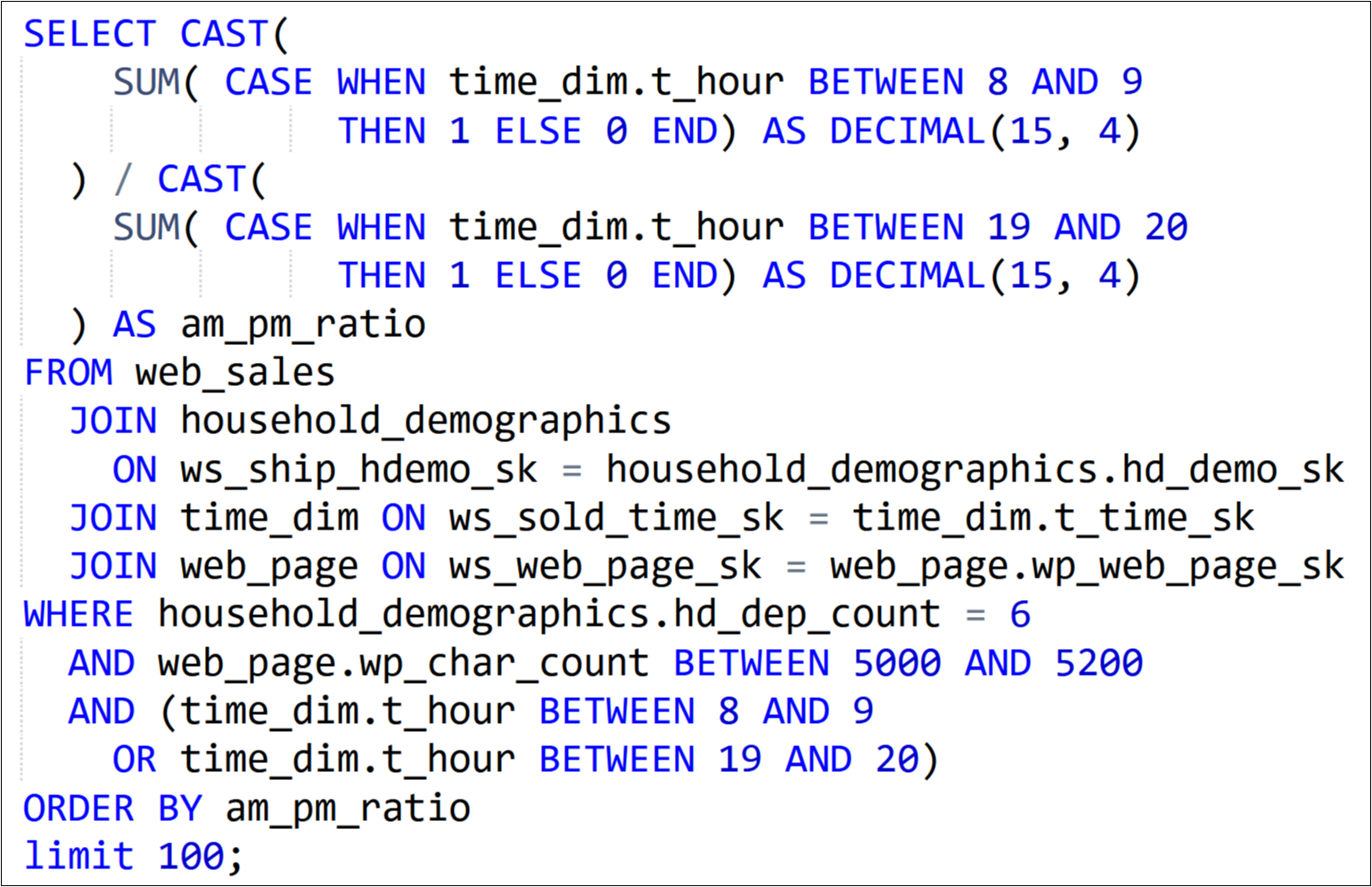}
    \vspace{-0.1in}
    \caption{Rewritten TPC-DS Q90}
    \label{fig:q90-rewrite}
    \vspace{-0.1in}
\end{figure}

On the flip side, there appears to be an ``impedance mismatch'' against such integration for certain classes of rewrites. 
For example, consider the TPC-DS Q90 rewrite in Figure~\ref{fig:q90-rewrite}. 
The original query individually computed  AM (morning) sales and PM (evening) sales, which were then used to compute the AM to PM ratio. The rewrite, however, 
extracted all relevant rows in one shot and computed the ratio using CASE statements -- encoding such transformations as generic rules in the optimizer appears challenging, given the combinatorial ways in which such transformations can occur.

Therefore, a fruitful area of future research could be achieving a middle-ground between the disparate world-views of LLMs and traditional optimizers.

\subsection{Agentic LLMs for Query Rewriting}

An effective way to extend \lithe is to use an agentic LLM that actively interacts with the database environment. 
This would allow a ``\lithe agent" to leverage database-related ``tools" such as the query optimizer's cardinality estimation and cost estimation modules, as well as rewrite-related tools such as rule generation, semantic validation, and regression checking. A memory store could be attached to the agent to log prior interactions with users and rewrite rules learnt over time, thus supporting continuous learning.
The agent can decide the order in which these tools are invoked and make appropriate choices during the rewrite process. 
Finally, combining our MCTS formulation with tool-augmented agentic reasoning would also be a promising avenue for future work.

\subsection{Scope of Semantic Equivalence Tools}
As seen in the experiments section, logic-based query equivalence testing covers industrial-strength queries only to a limited extent. 
On the other hand, while it is highly likely that the statistics-verified rewrites are valid, it still requires the DBA to make a final call on the correctness. 
This limitation restricts the use of \lithe in a fully automated scenario, i.e., as a direct preprocessor to the query engine. Therefore, a key challenge is to improve logic-based coverage. 

\subsection{Enhancing NL2SQL approaches with Rewrites}

\lithe could potentially be used to enhance NL2SQL approaches to produce performant SQL queries. One way would be to incorporate some rewrite rules directly into NL2SQL prompts. Based on our empirical assessment (Figure~\ref{fig:basic-prompts-cpr-breakup-tpcds}, we would recommend the simplest Prompt~1 from the Basic Prompts in conjunction with a subset of Rules~R1--R4. To avoid multiple prompts, one could use the LLM itself to identify the appropriate rule based on the NL query. While these prompts might not always provide the best rewrites, our experiments show that they are sufficient for many scenarios while incurring only small overheads as compared to using the entire prompt suite and MCTS search.  

An alternate non-invasive approach would be to simply use \lithe as a post-processor after the initial translation, but again restricted to using only a subset of the prompts/rules to keep the overheads within permissible limits.  

\section{Related Work}
\label{sec:Related-Work}

\myparagraph{Rule-based SQL rewriting.}
Most of the recent work on SQL query rewriting is rule-based~\cite{Learned_Rewrite, Wetune, QueryBooster, FactorWindow, Fusion, Starbust}. For instance, WeTune~\cite{Wetune} uses a rule generator to enumerate a set (up to a maximum size) of logically valid plans for a given query to create new rewrite rules, and uses an SMT solver to prove the correctness of the generated rules. While this approach can generate a large set of new rewrite rules, it often fails in coming up with transformation rules for complex queries due to the computational overheads of verifying rule correctness. As such, it is unable to rewrite any of the TPC-DS queries~\cite{Slabcity, Genrewrite}.

Learned Rewrite~\cite{Learned_Rewrite} uses existing Calcite~\cite{Calcite} rules and aims to learn the optimal subset of rules along with the order in which they must be applied. 
Since the rewrite search space grows exponentially with the number of rules, it uses MCTS scheme to efficiently navigate this space and find the rewritten query with maximum cost reduction.
This is in contrast to \lithe where MCTS is used to search over the output token space of an LLM in order to prevent LLM hallucinations. Specifically, our use of MCTS ensures that the LLM correctly follows the prompt instructions.

\llmrsq~\cite{LLMR2} is also rule-based but takes a different approach to identify the order for rewrite rule applications: it uses an LLM to find the best Calcite rules and the order in which to apply them to improve the query performance. R-Bot~\cite{rbot} also leverages an LLM to optimize the order of Calcite rules, but employs advanced contemporary techniques such as retrieval-augmented generation~(RAG) and step-by-step self-reflection to improve the outcomes. 

Query Booster~\cite{QueryBooster} implements human-centered rewriting -- it provides an interface to specify rules using an expressive rule language, which it generalizes to create rewrite rules to be applied on the query. 
There are also rule-based rewrite approaches designed for specific types of rewrites such as optimizing correlated window aggregations~\cite{FactorWindow} and common expression elimination~\cite{Fusion}. 

All of the above approaches operate via the query \emph{plan space}, which can restrict the kind of rewrites that can be accomplished. Whereas, \lithe uses a small set of general rewrite rules that work directly in the \emph{query space}.

\myparagraph{LLM-based rewriting.}
GenRewrite~\cite{Genrewrite} is the first LLM-based approach to use the LLM for end-to-end query rewriting.
Instead of using predefined rules from Calcite~\cite{Calcite}, they employ the LLM to create Natural Language Rewrite Rules (NLR2s) to be used as hints, and perform several iterations of prompting to get the rewritten query. 
They show that LLMs can outperform rule-based approaches due to their ability to understand contexts, and demonstrate a significant improvement in query rewriting compared to prior methods.

A limitation, however, is that LLM-generated rewrite rules often fail to generalize beyond specific query pairs. Even when generalized rules are present,  LLMs can struggle to apply rules correctly if not provided with accompanying examples. Finally, it must be noted that LLMs are unaware of the underlying database which restricts their ability to produce efficient metadata-aware rules. 

\myparagraph{Query rewriting in SQL engines.}
Query rewrite prior to plan enumeration has long been a feature of contemporary (non-LLM) database optimizers~\cite{Ahmed2006,Legaria2001}. However, the difference in our context is that the LLM gives us the flexibility to choose from a wide variety of approaches, not restricted to the hardwired options within individual systems.

\myparagraph{LLMs for Database Modules.}
LLM technologies have been advocated for a variety of database modules. For instance, they have been extensively used for Text-to-SQL transformations~\cite{BIRD, DB-GPT, sqleval, sqlcoder,2024CHESSpaper,DBGPT}.
Very recently, statistical metadata was leveraged to also improve the Text-to-SQL generation process~\cite{shkapenyuk2025}.
The main focus of these techniques is to correctly ascertain the information necessary to formulate the SQL query~\cite{MACSQL,DINSQL,2024CHESSpaper}.

On the other hand, the goal of SQL-to-SQL rewriting is on improving the performance of an existing SQL query.  Thus, unlike Text-to-SQL transformations where the input text is inherently ambiguous, SQL queries are precisely defined, and therefore
equivalence to a precise ground-truth has to be provably maintained. 

In recent times, LLMs have also been considered for a variety of database modules, including plan-hinting~\cite{LLMplanhint}, join-order optimization~\cite{LLMjoinorder}, index selection~\cite{DBGPT}, data pipelines~\cite{LLM_as_interface_for_data_pipeline}, data management~\cite{LLM_for_Data_Management}, and multi-modal data optimization~\cite{multi_model_query_optimizer}.  
These approaches can be used in conjunction with  \lithe since they address orthogonal segments of the query processing pipeline.

\section{Conclusions and Future Work}
\label{sec:Conclusion}

We investigated how the latent power of LLM technologies can be productively materialized in the context of SQL-to-SQL rewriting.  Our study progressively infused database domain knowledge, such as redundancy removal rules and schematic+statistical metadata, into the LLM prompts. Further, the output telemetry of LLMs, in the form of token probabilities, was used to signal situations where the LLM lacked confidence, triggering exploration of a larger search space. Finally, a combination of logic-based and statistical tests was employed to verify the equivalence of the rewrites.

An empirical evaluation over common database benchmarks showed that rewriting is a potent mechanism to improve query performance. In fact, even order-of-magnitude speedups were routinely achieved with regard to both abstract costing and execution times.
However, our results also showed a significant semantic distance between foundation models and query optimizers, with regard to both scope and precision, which would have to be bridged to fully leverage the latent power of LLMs. Further, our focus here was primarily on prompting-based strategies -- in our future research, we plan to investigate how domain-specific \emph{fine-tuning} could be leveraged to provide \gpt-like rewrites on small open models.

\newpage
\appendix
\begin{center}
    {\LARGE \bf Appendix}
\end{center}

\section{Additional Experiments}
\label{app:exp}

\subsection{Rewrite Quality (Other Benchmarks)}

We also evaluated \lithe on the following set of benchmarks. 
\begin{enumerate}
\item DSB~\cite{DSB}: A TPC-DS variant with complex data distributions and  additional query templates featuring many-to-many joins and non-equi-joins.
\item ARCHER~\cite{ARCHER}: A Text-to-SQL benchmark spanning 10 databases with availability of ground-truth SQL queries.
\item JOB~\cite{cardest}: An optimizer stress-test benchmark featuring queries with large and complex join graphs.
\item StackOverflow~\cite{Stackoverflow}: A real-world benchmark with query templates modeling questions and answers from experts. A random instance of each template is taken.
\end{enumerate}
Note that the queries in these workloads (with the exception of DSB) are mostly \textit{fast running} (i.e. completing in less than 10 seconds). Therefore, with the primary goal to test the coverage of \lithe over a diverse variety of workloads (and not the deployment angle), we include all the queries in our analysis.
The number of \cprs produced and the cost speedup delivered by \lithe and \sota over these benchmarks are shown in Table~\ref{tab:different_benchmark_peformance}.

\begin{table}[h]
\footnotesize
\centering
\caption{Comparing \lithe with {\sota} on different benchmarks.}
\label{tab:different_benchmark_peformance}
\begin{tabular}{|c|c|c|c|c|c|}
\hline
\multirow{2}{*}{\textbf{Benchmark}}
& \multicolumn{3}{c|}{\textbf{\# \cpr}} 
& \multicolumn{2}{c|}{\textbf{\csgm}} \\
& \textbf{LITHE} & \textbf{SOTA} & \textbf{Union}
& \textbf{LITHE} & \textbf{SOTA} \\ \hline \hline
DSB           & \LitheProdDSB           & \SotaProdDSB           & 9 & \gmLitheProdDSB           & \gmSotaProdDSB           \\ \hline
ARCHER      & \LitheProdArcher        & \SotaProdArcher        & 22 & \gmLitheProdArcher        & \gmSotaProdArcher        \\ \hline
JOB          & \LitheProdJOB           & \SotaProdJOB           & 4 & \gmLitheProdJOB           & \gmSotaProdJOB           \\ \hline
StackOverflow & \LitheProdStackoverflow & \SotaProdStackoverflow & 2 & \gmLitheProdStackoverflow & \gmSotaProdStackoverflow \\ \hline
\end{tabular}
\end{table}

In case of DSB, \lithe produces \cpr for 9 queries resulting in a highly productive \csgm of \textbf{7.7}. Similar to the case with TPC-DS, \lithe performs better than \sota both with respect to its coverage, as well as the cost speedups.

Turning our attention to the other benchmarks (ARCHER, JOB, StackOverflow), the number of \cpr queries is smaller due to the predominance of flat SPJ formulations in these benchmarks, which limits the scope for productive rewriting. 
Nevertheless,  \lithe continues to achieve better \cpr coverage, whereas \sota misses quite a few opportunities. Further, the \csgm of \lithe is visibly better than \sota.

\subsection{Rewrite Quality (Rule-based vs LLM-based approaches)}

Our \sota baseline comprises two theoretical rule-based approaches—Learned Rewrite and LLM-R2—and two LLM-based approaches—GenRewrite and Baseline Prompt. The detailed breakdown of CPR achieved by these methods is presented in Table~\ref{tab:cpr-breaup-sota}.

\begin{table}[!h]
\footnotesize
\centering
\caption{\# CPR of Different SOTA Approaches in Comparison with LITHE}
\label{tab:cpr-breaup-sota}
\begin{tabular}{|c|c|c|c|}
\hline
\textbf{Tools/Techniques} & \textbf{Approach}&\textbf{Individual \# \cpr}& \textbf{$>=$1.1x} \\ 
\hline \hline
\lithe & LLM Based & 26 & 30 \\ \hline\hline
Genrewrite & LLM Based & 11 & 12\\ \hline
Baseline & LLM Based & 10 & 10\\ \hline
Learned Rewrite & Rule Based & 1 & 2 \\ \hline
LLMR2 & Rule Based & 0 & 0 \\\hline
\end{tabular}
\vspace{-0.1in}
\end{table}

While the theoretical rule-based approaches provide formal correctness guarantees, our TPC-DS experiments revealed that the LLM-based approaches consistently outperformed them. Among the 13 CPRs (i.e., query rewrites with cost speedup $\geq$ 1.5X), only one was produced by a theoretical rule-based approach. Moreover, even when we decreased this threshold to as low as $\geq$ 1.1X,  only one more rewrite became a CPR!

LLM-based approaches, on the other hand, seem more suited for query rewrites, resulting in 12 (GenRewrite) and 30 (\lithe) rewrites with a cost speedup $\geq$ 1.1X.
This is likely a consequence of the LLM having a ``global'' vision of the entire query, giving it more leeway in considering rewrites in the query-space. Whereas, theoretical rule-based approaches operate mainly at the operator level within the plan tree, thus being confined to only local optimizations.

\subsection{Rewrite quality (Coverage of Basic Prompts)}
\label{app:basic-prompts-rewritten-queries-overlap}

\begin{figure}[h]
    \centering
    \begin{subfigure}[b]{0.49\linewidth}
    \centering
    \includegraphics[width=\linewidth]{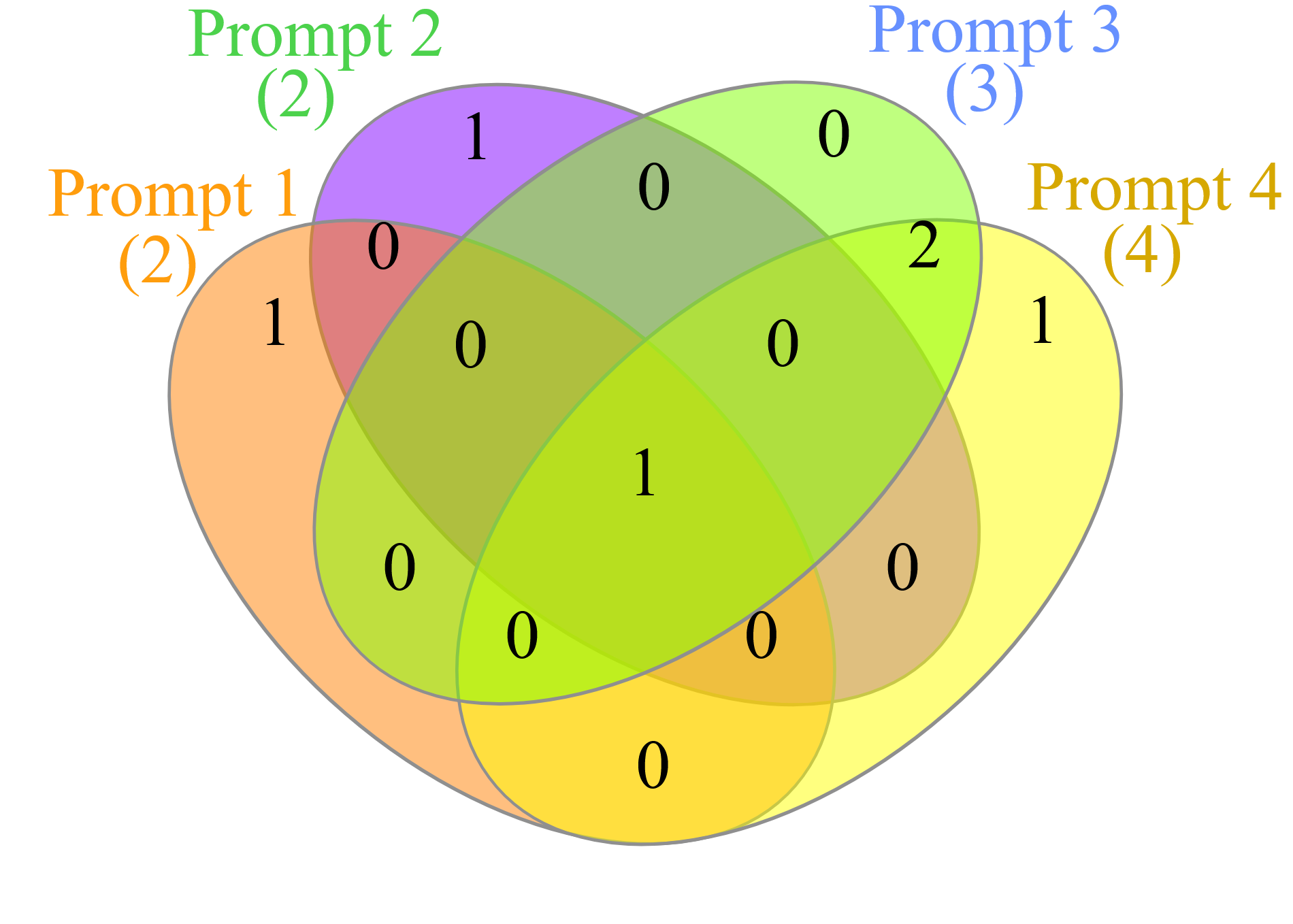}
    \caption{Micro-benchmark}
    \label{fig:basic-prompts-cpr-breakup-microbenchmark}
    \end{subfigure}
    \begin{subfigure}[b]{0.49\linewidth}
    \centering
    \includegraphics[width=\linewidth]{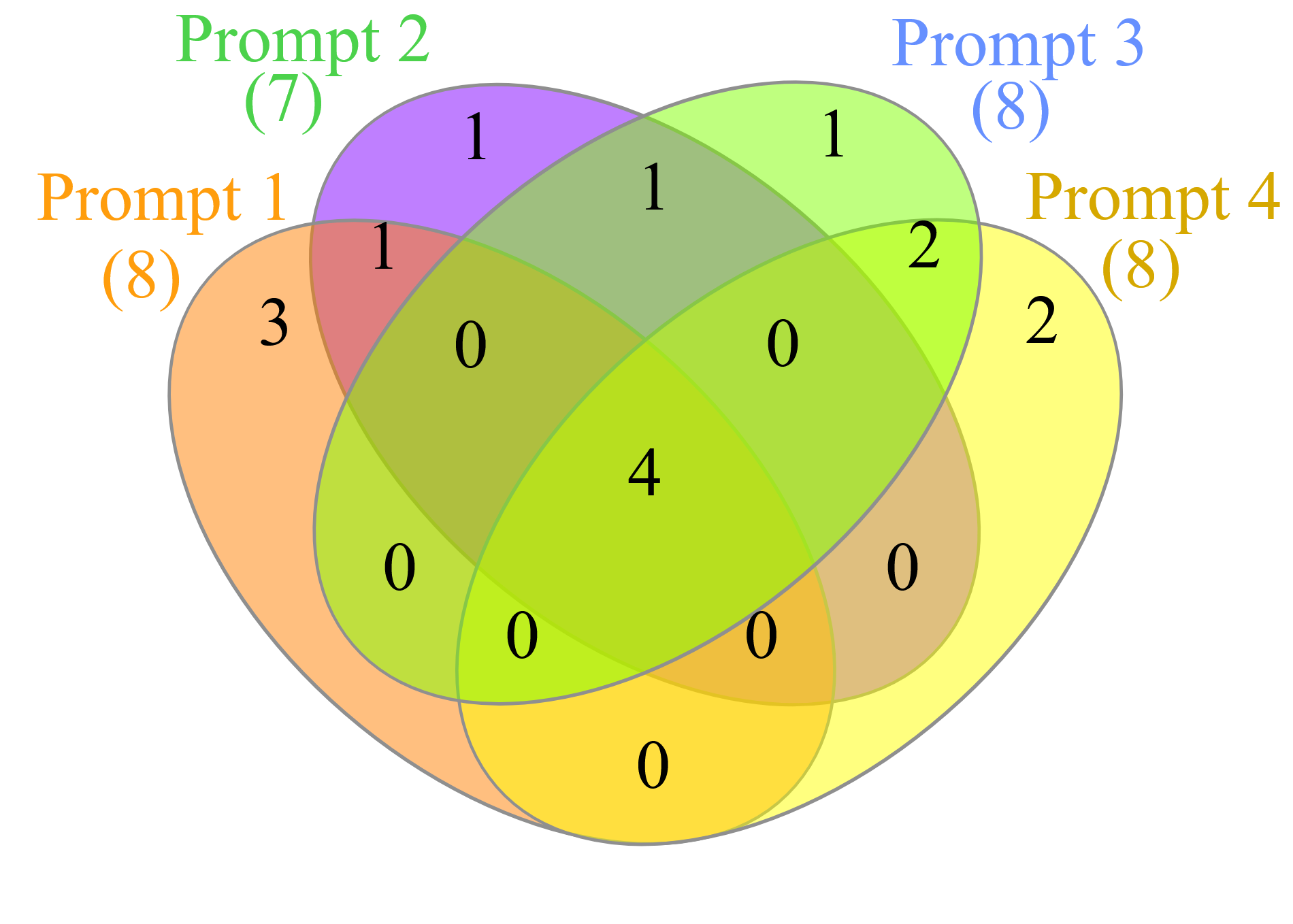}
    \caption{TPC-DS}
    \label{fig:basic-prompts-cpr-breakup-tpcds}
    \end{subfigure}
    \caption{CPR distribution of Basic Prompts}
\end{figure}

Figure \ref{fig:basic-prompts-cpr-breakup-microbenchmark} shows the CPR distribution across the Basic Prompts on the Microbenchmark. Among the 6 rewritten queries, 3 are unique to a single prompt, while the other 3 are produced by multiple prompts. At first glance,
one might think that Prompt 3 could be removed from our prompt suite since it does not produce any unique rewrite. However, this is an artifact of the micro-benchmark -- on the corresponding picture for the full TPC-DS benchmark (Figure \ref{fig:basic-prompts-cpr-breakup-tpcds}), Prompt 3 does contribute unique rewrites. 

\newpage
\section{Overheads Reduction for \lithe}
\label{app:classifier}

\subsubsection*{Rule Selection Classifier}
Reducing the query rewrite time while still obtaining the same performance would be possible if we could directly use the MCTS-driven rewrite with the appropriate prompt. Towards this end, we build a classifier to pick the most appropriate rewrite rule for a given input query.
Specifically, the classifier identifies which, if any, of Rules R1--R6 is appropriate for a given query. If none are appropriate, then it falls back to just the set of basic prompts to identify the best prompt to be given as input to the MCTS module.
In addition to reducing the rewrite times, using a classifier can also reduce the financial costs of the rewrite.

We design an LLM based classifier to accomplish this as follows: The LLM is given the rewrite rules discussed so far, and additionally, for each rule, an example demonstrating when the rule can be applied and a counter-example demonstrating when the rule cannot be applied. For the database schema and statistics-based rules, the relevant information is also fed to the classifier so that it can make an informed decision. Then, given an input query, the classifier is tasked with selecting the most appropriate rewrite rule.

\begin{table}[h]
\footnotesize
\centering
\caption{Impact of Classifier (TPC-DS)}
\label{tab:classifier-tpcds}
\begin{tabular}{|c|c|c|}
\hline
\textbf{Metrics} & \textbf{Without Classifier} & \textbf{With Classifier} \\ \hline \hline
\# \cpr & 26 & 23 \\ \hline
\csgm & 11.5 & 8.5 \\ \hline
Avg. Tokens & 18427 & 16003 \\ \hline
Avg. Time & 5~min & 2.4~min \\ \hline
\end{tabular}
\end{table}

Table~\ref{tab:classifier-tpcds} compares the performance of \lithe with and without the classifier. The time overheads do visibly go down by about 52 percent, and the tokens by about 13 percent. However, there is a price to be paid -- the \cpr is reduced to 23 and the \csgm comes down to 8.5. In our future work, we plan to look into whether a better tradeoff could be achieved between quality and overheads.

As an alternative to the LLM based classifier, we also experimented with a DSPy MIPRO \cite{MIPRO} based classifier to identify which, if any, of Rules R1–R6 is appropriate for a given query. However, our initial results indicated that it was weaker than our LLM-based classifier. While the average time taken by this approach reduced to just 10 seconds per query, the \cprs are reduced to 14, and the \csgm and \tsgm fall down to 3.3 and 4.5 respectively.

But we hasten to add that this outcome is likely an artifact of our small rule set – looking to the future, where there may be a larger corpus, a data-driven selection method like MIPRO may come to the fore.

\subsubsection*{Pruning in MCTS}
A bottleneck in the MCTS-based exploration is the need to greedily expand a node (during the simulation stage) until an entire rewrite is output. In principle, if we could quickly check for semantic and syntactic correctness at intermediate stages, then unproductive paths could be terminated early. We are currently working on the design and implementation of such checks.

\newpage
\section{Query Equivalence Testing via Sampling}
\label{app:sampling-eq}

We use a sampling-based approach to quickly test equivalence in the rewrite generation stages of the pipeline. The idea here is to execute the queries on several small samples of the database and verify equivalence based on the sample results. 
However, while this test is a necessary condition for query equivalence, it is not a sufficient condition. That is, there are no false negatives, but there can be false positives. This is because the sampled database may not cover all the predicates present in the query. This can cause two types of problems: 
First, it is possible for two different queries to return the same result. This can happen when, for example, the entirety of the sampled data satisfies a predicate of one query, while the same predicate is not present in the other. Second, if the underlying sample does not satisfy any of the predicates in either query, then an empty result will be returned by both queries. This again does not imply that the queries are equivalent.

To statistically address the first problem (false positives), we create multiple samples of the database with different seeds, and run the test on all these samples. The goal is to reduce the likelihood of non-equivalent queries returning the same results. 

To minimize the occurrence of the second problem (empty results), the following approach is taken:
\begin{enumerate} \denselist
\item We use \textit{correlated sampling}~\cite{cs2} to sample the database. This technique leverages the join graph of the schema to produce a sample that maintains join integrity between the tables participating in the query.
\item Given a pair of queries to test for equivalence, we adjust the constants in the filter predicates to reduce the chances of an empty result. We make use of the rows in the sampled data for this purpose. For example, say an equality predicate is present in the query and the associated constant is absent in the sampled database. We then replace the query constant with a value already present in the sample. Similarly, the constants for other comparison operators are adjusted based on the ranges of the corresponding columns in the sampled database. 

Note that these modifications only change the selectivity of the query, but not its semantics.

\item  To cover predicate boundary conditions, we inject query-specific synthetic tuples into the sample—similar to the approach used by the XData mutant-killing tool~\cite{xdata}—in order to reduce the likelihood of non-equivalent queries producing the same results.

\end{enumerate}

\newpage

\begin{flushleft}
\section{Examples used in Prompts for Rules 1--6}
\label{app:rules-ex}

\vspace{0.1in}
\noindent \textcolor{blue}{\large \textbf{R1: Use CTEs (Common Table Expressions) to avoid repeated computation.}}

\vspace{0.2in}
\myparagraph{\underline{Original Query}}

        \begin{verbatim}
SELECT emp.employee_name,
       mgr.manager_name
FROM   employees emp,
       managers mgr
WHERE  emp.manager_id = mgr.manager_id
       AND emp.employee_id IN (SELECT manager_id
                               FROM   (SELECT manager_id,
                                              manager_name
                                       FROM   managers
                                       WHERE  job_id = 'IT_PROG'
                                              AND manager_id > 100))
       AND mgr.manager_name IN (SELECT manager_name
                                FROM   (SELECT manager_id,
                                               manager_name
                                        FROM   managers
                                        WHERE  job_id = 'IT_PROG'
                                               AND manager_id > 100)); 
\end{verbatim}

\myparagraph{\underline{Rewritten Query}}

\begin{verbatim}
WITH cte
     AS (SELECT manager_id,
                manager_name
         FROM   managers
         WHERE  job_id = 'IT_PROG'
                AND manager_id > 100)
SELECT emp.employee_name,
       mgr.manager_name
FROM   employees emp,
       managers mgr
WHERE  emp.manager_id = mgr.manager_id
       AND emp.employee_id IN (SELECT manager_id
                               FROM   it_prog_managers)
       AND mgr.manager_name IN (SELECT manager_name
                                FROM   it_prog_managers); 
        \end{verbatim}

\newpage
\noindent \textcolor{blue}{\large \textbf{R2: When multiple subqueries use the same base table, rewrite to scan the base table only once.}}

\vspace{0.2in}
\myparagraph{\underline{Original Query}}
        \begin{verbatim}
SELECT (SELECT Avg(salary)
        FROM   employees
        WHERE  department = 'Sales'
               AND experience_years BETWEEN 1 AND 5
               AND salary BETWEEN 50000 AND 60000) AS Sales_Avg,
       (SELECT Avg(salary)
        FROM   employees
        WHERE  department = 'HR'
               AND experience_years BETWEEN 5 AND 10
               AND salary BETWEEN 80000 AND 90000) AS HR_Avg; \end{verbatim}
\vspace{0.2in}
\myparagraph{\underline{Rewritten Query}}
\begin{verbatim}
SELECT avg(
       CASE
              WHEN department = 'Sales' THEN salary) AS sales_avg,
       avg(
       CASE
              WHEN department = 'HR' THEN salary) AS hr_avg
FROM   employees
WHERE  (
              department = 'Sales'
       AND    experience_years BETWEEN 1 AND    5
       AND    salary BETWEEN 50000 AND    60000)
OR     (
              department = 'HR'
       AND    experience_years BETWEEN 5 AND    10
       AND    salary BETWEEN 80000 AND    90000);
       \end{verbatim}

\newpage
\noindent \textcolor{blue}{\large \textbf{R3: Eliminate overlapping subqueries.}}

\vspace{0.2in}
\myparagraph{\underline{Original Query}}
        \begin{verbatim}
SELECT c.*
FROM   customer c
WHERE  c.address_id IN (SELECT a.address_id
                        FROM   address)
       AND c.address_id IN (SELECT a.address_id
                            FROM   address
                            WHERE  a.pin_code = '560012'); \end{verbatim}
\vspace{0.3in}
\myparagraph{\underline{Rewritten Query}}
\begin{verbatim}
SELECT c.*
FROM   customer c
WHERE  c.address_id IN (SELECT a.address_id
                        FROM   address
                        WHERE  a.pin_code = '560012'); \end{verbatim}

\newpage
\noindent \textcolor{blue}{\large \textbf{R4: Remove unnecessary joins between a primary key and a foreign key.}}

\vspace{0.2in}
\myparagraph{\underline{Schema}}
        \begin{verbatim}
CREATE TABLE products
  (
     p_product_id INTEGER NOT NULL,
     PRIMARY KEY (p_product_id)
  );

CREATE TABLE fact_sales
  (
     f_sales_id   INTEGER NOT NULL,
     f_units_sold INTEGER NOT NULL,
     f_product_id INTEGER NOT NULL,
     PRIMARY KEY (f_sales_id),
     FOREIGN KEY (f_product_id) REFERENCES products(p_product_id)
  ); 
  
  \end{verbatim}

        \item \textbf{\underline{Original Query}}
        \begin{verbatim}
SELECT p_product_id,
       f_units_sold
FROM   fact_sales,
       products
WHERE  f_product_id = p_product_id; \end{verbatim}

\vspace{0.2in}
\myparagraph{\underline{Rewritten Query}}
\begin{verbatim}
SELECT f_product_id,
       f_units_sold
FROM   fact_sales; 

\end{verbatim}
\newpage
\noindent \textcolor{blue}{\large \textbf{R5: Choose EXIST or IN based on subquery selectivity.}}

\vspace{0.2in}
\myparagraph{\underline{Original Query}}
        \begin{verbatim}
Select item.id, item.code, item.price 
from item 
where item.sourceid in (
    Select element.sourceid 
    from element 
    where element.zip > 1100
  ) 
order by item.id;\end{verbatim}

\vspace{0.1in}
\myparagraph{\underline{Statistics}}
    \begin{verbatim}
Selectivity of different predicates is given below : 
( 1 ) source_id > 1100 on table element :: 0.7385
    \end{verbatim}
    
\vspace{0.1in}
\myparagraph{\underline{Rewritten Query}}
\begin{verbatim}
Select item.id, item.code, item.price 
from item 
where exists(select 1 
    from element 
    where item.sourceid = element.sourceid 
      and element.sourceid > 1100
  ) 
order by item.id;
 \end{verbatim}
 
\newpage
\noindent\textcolor{blue}{\large \textbf{R6: Pre-filter tables that are involved in self-joins and have low selectivities on their filter and/or join predicates. Remove any redundant filters from the main query. Do not create explicit join statements.}}

\vspace{0.2in}
\myparagraph{\underline{Original Query}}
\begin{verbatim}
with total_price_cte as (  
    select item.id, colour.colorcode, sum(item.price) total_price  
    from item, color  
    where item.colorcode = colour.colorcode  
    group by item.id, colour.colorcode  
)  
select t_sec.id, t_first.colorcode  
from total_price_cte t_first, total_price_cte t_sec  
where t_sec.id = t_first.id  
    and t_first.colorcode = `R'  
    and t_sec.colorcode = `R'  
    and t_first.total_price > 0  
order by t_sec.id  
limit 100;
\end{verbatim}

\vspace{0.1in}
\myparagraph{\underline{Statistics}}
\begin{verbatim}
Selectivity of different predicates is given below : 
( 1 ) colour.colorcode = `R' :: 0.01
\end{verbatim}

\vspace{0.1in}
\myparagraph{\underline{Rewritten Query}}
\begin{verbatim}
with total_price_cte as (  
    select item.id, colour.colorcode, sum(item.price) total_price  
    from item, color  
    where item.colorcode = colour.colorcode  
    group by item.id, colour.colorcode  
),  
filtered_total_price_cte as (  
    select * from total_price_cte  
    where colorcode = `R'  
)  
select t_sec.id, t_first.colorcode  
from filtered_total_price_cte t_first, filtered_total_price_cte t_sec  
where t_sec.id = t_first.id  
    and t_first.total_price > 0  
order by t_sec.id  
limit 100;
 \end{verbatim}

\end{flushleft}

\end{document}